\documentclass[iop]{emulateapj}

\usepackage{bm}
\usepackage{float}
\usepackage{amsmath}
\usepackage{mathrsfs}

\usepackage{upgreek}
\usepackage{graphicx}
\usepackage{color}

\newcommand{\msunyr}{M_{\odot}\:{\rm{yr}}^{-1}}
\newcommand{\mdot}{\dot{m}}
\newcommand{\Mdot}{\dot{M}}
\newcommand{\msun}{M_{\odot}}
\newcommand{\rsun}{R_{\odot}}
\newcommand{\lsun}{L_{\odot}}
\newcommand{\zsun}{Z_{\odot}}

\newcommand{\HII}{\ion{H}{2}}
\newcommand{\Sigcl}{\Sigma_{\rm{cl}}}
\newcommand{\gcm}{{\rm g\:cm^{-2}}}
\newcommand{\kms}{{\rm\:km\:s^{-1}}}

\newcommand{\sfe}{\bar{\varepsilon}_{\rm{*f}}}
\newcommand{\N}{\mathscr{N}}

\renewcommand{\thefootnote}{\fnsymbol{footnote}}

\shorttitle{Feedback During Massive Star Formation}
\shortauthors{Tanaka, Tan \& Zhang}

\begin{document}

\title{The Impact of Feedback During Massive Star Formation by Core Accretion}

\author{Kei E. I. Tanaka}
\affil{Department of Astronomy, University of Florida, Gainesville, FL 32611, USA; ktanaka@ufl.edu}
\author{Jonathan C. Tan}
\affil{Departments of Astronomy \& Physics, University of Florida, Gainesville, FL 32611, USA; jctan.astro@gmail.com}
\and
\author{Yichen Zhang}
\affil{Departamento de Astronom\'ia, Universidad de Chile, Santiago, Chile}
\affil{Star and Planet Formation Laboratory, RIKEN, Wakoshi, Saitama, 351-0198, Japan; yczhang.astro@gmail.com}

\begin{abstract}
We study feedback during massive star formation using semi-analytic
methods, considering the effects of disk winds, radiation pressure,
photoevaporation and stellar winds, while following protostellar
evolution in collapsing massive gas cores.  We find that disk winds
are the dominant feedback mechanism setting star formation
efficiencies (SFEs) from initial cores of $\sim0.3$--$0.5$.  However,
radiation pressure is also significant to widen the outflow cavity
causing reductions of SFE compared to the disk-wind only case,
especially for $>100\msun$ star formation at clump mass surface
densities $\Sigcl\lesssim0.3\:\gcm$.  Photoevaporation is of
relatively minor importance due to dust attenuation of ionizing
photons.  Stellar winds have even smaller effects during the accretion
stage.  For core masses $M_c\simeq10$--$1000\:\msun$ and
$\Sigcl\simeq0.1$--$3\:\gcm$, we find the overall SFE to be
$\sfe=0.31(R_{c}/0.1\:{\rm{pc}})^{-0.39}$, potentially a useful
sub-grid star-formation model in simulations that can resolve
pre-stellar core radii,
$R_c=0.057(M_c/60\msun)^{1/2}(\Sigcl/\gcm)^{-1/2}\:{\rm{pc}}$.  The
decline of SFE with $M_c$ is gradual with no evidence for a maximum
stellar-mass set by feedback processes up to stellar masses of
$m_*\sim300\:\msun$.  We thus conclude that the observed truncation of
the high-mass end of the IMF is shaped mostly by the pre-stellar core
mass function or internal stellar processes.  To form massive stars
with the observed maximum masses of $\sim150$--$300\msun$, initial
core masses need to be $\ga500$--$1000\:\msun$.  We also apply our
feedback model to zero-metallicity primordial star formation, showing
that, in the absence of dust, photoevaporation staunches accretion at
$\sim50\:\msun$.  Our model implies radiative feedback is most
significant at metallicities $\sim10^{-2}\zsun$, since both radiation
pressure and photoevaporation are effective in this regime.
\end{abstract}


%%%%%%%%%% Section 1 %%%%%%%%%%
\section{Introduction}\label{sec_intro}

Massive stars play important roles in a wide range of astrophysical
settings. They are the sources of UV radiation, turbulent energy and
heavy elements. Massive star close binaries are the likely progenitors
of merging black hole systems that have been the first sources to be
detected by their gravitational wave emission. However, massive star
formation is relatively poorly understood compared to low-mass star
formation \citep[see][for a recent review]{tan14}. One class of models
of massive star formation is based on the Core Accretion scenario
\citep[e.g., the Turbulent Core Model of][]{mck03}. These models are
scaled-up versions of models of low-mass star formation from cores
(e.g., Shu, Adams \& Lizano 1987), invoking nonthermal forms of
pressure support, i.e., turbulence and magnetic fields to help
stabilize the initial massive pre-stellar core. However, there may
also be significant differences compared to low-mass star formation
due to the stronger feedback that is expected from massive protostars.

In low-mass star formation, the magnetohydrodynamic (MHD) outflow is
thought to be the main feedback process, which may determine the star
formation efficiency (SFE) from the pre-stellar core, i.e.,
$\sfe\equiv m_{*f}/M_{c}$ where $m_{*f}$ is the final mass that is
achieved by the protostar at the end of its accretion and $M_c$ is the
mass of the initial core.
In relatively low-mass clusters that contain stars with masses up to
$\sim10\msun$, the core mass function (CMF) is reported to be similar
in shape to the stellar initial mass function (IMF), but shifted to
higher masses by a factor of a few \citep[e.g.,][]{and10,kon10}.
One explanation for this is a nearly constant SFE as a function of
core mass of about $\sfe\sim0.4$. \citet{mat00} proposed that the
accretion-powered, MHD-driven outflow sets the SFE from pre-stellar
cores. They provided an analytic model and showed that the momentum
injected by the disk wind sweeps up a certain fraction of material in
the infalling envelope and sets a SFE of $\sim0.3$--$0.5$.  The
numerical simulation by \citet{mac12} confirmed this result obtaining
a similar value of SFE.  Therefore, in low-mass star formation,
observations and theoretical models are in agreement that an
individual star can be formed by collapse of a pre-stellar core with
the MHD outflow setting a SFE of $\sim0.4$.

In massive star formation, additional feedback processes may become
more significant than the MHD outflow because of the high luminosities
of massive stars.  Especially, radiation pressure has been considered
to be a potential barrier for massive star formation.  In an idealized
spherical geometry, radiation pressure acting on dust grains in an
infalling envelope exceeds the gravitational force when the stellar
mass reaches $\sim10$--$20\:\msun$ preventing further mass accretion
\citep{lar71,wol87}. The fact that more massive stars exist tells us
that the model of spherical infall is too simplified. Subsequent work
on analytic and semi-analytic models
\citep[e.g.,][]{nak89,jij96,tan11} and numerical simulations
\citep[e.g.,][]{yor02,kru09,kui10,ros16} of disk accretion found that
mass infall and accretion can continue from behind the disk since this
region is shielded from strong radiation pressure.  The series of
simulations by Kuiper and collaborators have shown that disk accretion
continues while the direct stellar radiation sweeps up the material
above the disk where the shielding effect is weak
\citep{kui10,kui11,kui12,kui15,kui16}.  They found the SFE from
$100\msun$-cores is about $0.5$ in models without MHD disk wind
feedback. The recent simulation with high resolution and moving sink
particle method by \citet{ros16} showed that the Rayleigh-Taylor (RT)
instability strongly helps to bypass the radiation pressure barrier
even above the disk.  Also, if MHD outflow cavities exist before
radiation pressure becomes significant, then radiation leaks away via
these channels, i.e., enhancing the so-called ``flashlight effect''
\citep{yor99,yor02,kru05,kui15,kui16}. Thus, the radiation pressure
barrier is not thought to be a catastrophic problem anymore for
massive star formation. Rather the question now is what is its
quantitative effect on the formation efficiency of massive stars from
massive cores.

Photoionization may also be a significant feedback process. When a
massive protostar approaches the Zero-Age Main-Sequence (ZAMS), it
contracts, increases its effective temperature and starts to emits
significant fluxes of Lyman continnum photons with $>13.6{\rm eV}$
that may ionize the infalling/accreting material. Such ionized gas has
a high temperature of $\sim10^4\:{\rm K}$ and its thermal pressure may
drive mass-loss in a ``photoevaporative'' outflow. In the formation of
primordial (Pop III) stars in the early universe, photoevaporation is
thought to be significant, potentially stopping mass accretion at
$\sim50$--$100\msun$ \citep{mck08,hos11,tan13}. Note that radiation
pressure feedback is not very significant in primordial star formation
since there are no dust grains. Coincidentally, the typical mass
accretion rates in primordial star formation and in present-day
massive star formation are expected to be similar, with values of
$\sim10^{-3}\:\msunyr$.  Thus, one may speculate that photoevaporation
also stops mass accretion in present-day massive star formation.
However, the dependence of the photoevaporation rate on metallicity
has not been studied very much and remains uncertain. The simulation
of present-day massive star formation by \citet{pet10} suggested that
photoionization feedback is not very significant, but a general
theoretical framework of photoevaporation spanning the whole
metallicity range from primordial to quasi-solar metallicities remains
lacking.

Feedback by protostellar outflows, radiation pressure and
photoevaporation act on the infalling/accreting material. Stellar
winds launched from the protostellar surface could in principle also
act against the accretion flow, but, as we will discuss below, they
are expected to always be confined by the protostellar outflow and
thus not have a direct impact on the accretion.
However, mass-loss directly from these stellar winds could potentially
become significant, especially for protostars at the highest masses
and luminosities. The mass-loss by a stellar wind is certainly
important during the later evolution of massive stars.  For example,
in the case of the R136a1 Wolf-Rayet (WR) star with current mass of
$265\msun$, a stellar wind mass-loss rate of $5\times10^{-5}\msunyr$
is inferred, and its initial mass is evaluated to have been as high as
$320\msun$ \citep{cro10}.  The theoretical calculation by
\citet{vin11} has found that the stellar wind mass-loss rate becomes
extremely high if the Eddington factor to electron scattering is
higher than $0.7$.  They interpreted this high mass-loss regime as
leading to the observational appearance as WR stars and the lower
mass-loss regime as O-type stars.
However, the protostellar internal luminosity as a function of mass,
and thus the Eddington factor, depends on the accretion history. Thus
it is possible that in some circumstances the Eddington factor might
potentially reach the extreme mass-loss regime even during the
protostellar stage.

The feedback and mass-loss processes described above may impact the
ability of very massive stars to form and thus reveal themselves in
the observed distribution of the IMF, e.g., perhaps creating a break
or turnover in the \citet{sal55} power law that holds from lower
masses $\sim1\:M_\odot$ to at least $\sim 100\:M_\odot$. In other
words, feedback and mass-loss may imply there is a maximum stellar
mass that can form.
Observationally, \citet{fig05} have reported the absence of stars with
masses $>150\:\msun$ in the Arches cluster near the Galactic center,
whereas extension of the Salpeter mass function predicts there should
be 18 of them. Thus Figer concluded there is an upper stellar mass
limit of $150\:\msun$. Later studies of the Tarantula nebula in the
Large Magellanic Cloud (LMC), stars with initial masses of
$200$--$300\:\msun$ were found \citep{cro10,bes11}. Since LMC has
lower metallicity (by about a factor of two) than the Galaxy,
it can be speculated that the impact of feedback and/or mass-loss
depends on the metallicity, which then affects the upper IMF in
different environments. 

A trend to a higher maximum stellar mass with decreasing metallicity
is potentially supported by the theoretical studies of
\citet{hir14,hos16}, who found that Pop III stars may reach masses as
high as $1000\:M_\odot$. Unfortunately there are no direct
observational constraints on the masses of Pop III stars.  However,
the chemical abundance patterns of Galactic metal-poor stars, which
may be second generation stars polluted by Pop III supernova ejecta,
have been interpreted as indicating that there were such very massive
primordial stars \citep{kel14,aok14}. Such conclusions, however,
remain very tentative. Overall, a good theoretical understanding of
how the stellar IMF depends on metallicity remains lacking.

Although there have been many studies concentrating on each radiative
feedback mechanism in massive star formation, there has not yet been a
study that has considered all the main processes together, including
with the effects of an MHD-launched outflow. In this paper, we aim to
carry out such a study with the goal of evaluating the SFE of
pre-stellar cores of different masses and in different environments. A
full numerical simulation with MHD and radiative effects that resolves
the protostellar surface and the outer core scale and follows the full
evolutionary growth of the protostar is computationally challenging
and beyond current state-of-the-art capabilities. Here we present a
semi-analytic model of this process that includes all the expected
important physical processes and yet at the same time can be applied
to large range of different conditions. This allows us to gain
physical insight into the problem and can help guide future numerical
simulation experiments. Our modeling builds upon our previous work
that developed semi-analytic models for massive star formation
\citep{zha11,zha13,zha14,tan16}, but which did not yet include
treatment of radiative feedback or stellar wind
mass-loss. Additionally, we apply the same model to primordial star
formation at zero metallicity, to gain insight into the metallicity
dependence of massive star formation feedback.

This paper is organized as follows. In \S\ref{sec_method} we review
the basics of our model, and introduce the updates to include the
effects of feedback processes.  Next, in \S\ref{sec_results},
we present our results:
we show how the accretion rate and SFE are reduced
by multiple feedback processes, and also reveal the differences
caused by solar and zero metallicities.
In \S\ref{sec_discussion}, we discuss the relative importance of
different feedback mechanisms, their impact on shaping the high-mass
end of the IMF, and their dependence on metallicity.  We conclude in
\S\ref{sec_conclusion}.

%%%%%%%%%% Section 2 %%%%%%%%%%
\section{Methods}\label{sec_method}

We calculate the accretion history of massive star formation including
multiple feedback processes. The framework of our model has been
constructed in a series of papers: \citet{zha11,zha13,zha14}, and
\citet{tan16}. In these works, massive protostellar evolution with MHD
disk wind feedback is calculated. We then estimate continuum emission
from the protostar and disk, which is then followed in radiative
transfer calculations, especially to predict infrared, sub-millimeter
and cm-radio morphologies and spectral energy distributions.
Now we extend this massive protostellar evolution model to include
feedback by radiation pressure and photoevaporation, and stellar wind
mass-loss.  We note that this follows a similar methodology to that of
\citet{mck08}, who considered the formation of primordial stars under
the influence of multiple feedback processes.

\begin{figure}
\begin{center}
\includegraphics[width=85mm]{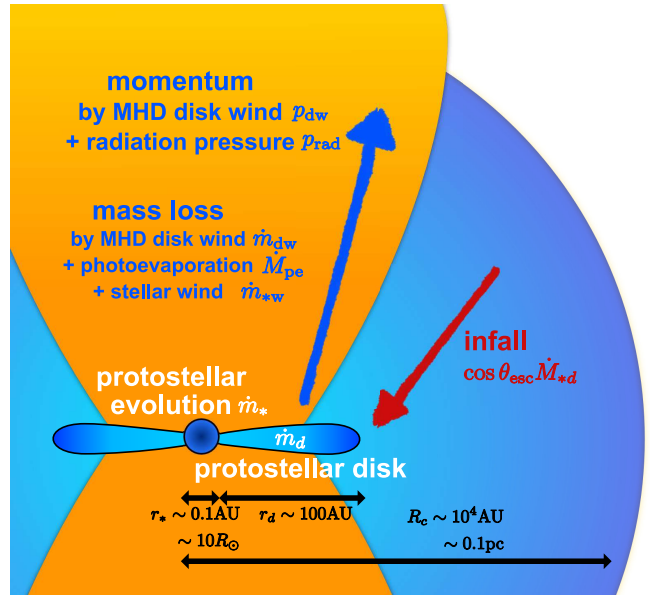}
\end{center}
\caption{
Schematic view of massive star formation by core accretion including
various feedback processes. The parameters for the initial conditions
are core mass $M_c$, mass surface density of the ambient clump
$\Sigcl$, and the ratio of core initial rotational to gravitational
energy $\beta_{\rm c}$. The model includes momentum feedback from a
MHD disk wind and radiation pressure. It also follows mass-loss
resulting from the MHD disk wind, photoevaporation and the stellar
wind.}
\label{fig_schematic}
\end{figure}

We review the basics of our model that were developed in previous
works \citep{zha11,zha13,zha14} in \S\ref{sec_method_rev}, introduce
the methods for each feedback process in \S\ref{sec_method_fb}, and
explain how they are combined together in \S\ref{sec_acc_rate}.
Figure \ref{fig_schematic} shows the schematic view of our model.
Although the main target of this study is present-day massive star
formation, we also apply our model to primordial star formation for
comparison of this different environment and for comparison with the
previous results of McKee \& Tan (2008). Thus in
\S\ref{sec_popIII_model}, we describe the modifications of methods
that are used to apply the model to primordial star formation.

\subsection{Evolution of infall rates, disks and protostars}\label{sec_method_rev}

Our model assumes a pre-stellar core collapses to form one massive
star. This model should be a reasonable approximation even for
multiple systems in which there is a single dominant protostar. The
initial core is assumed to be spherical and close to virial
equilibrium by the support of turbulence and/or magnetic fields
\citep{mck03}. The parameters to determine core properties are core
mass $M_c$, mass surface density of the ambient clump $\Sigcl$,
and the core's initial rotational to gravitational energy ratio
$\beta_c$.
The core is assumed to be in pressure equilibrium with the ambient
clump. If the clump is self-gravitating then this ambient pressure is
related to its surface density $\Sigcl$, which sets the pressure at
the core surface, thus determining its size.  The core radial density
profile is assumed to be a power law, i.e., $\rho\propto
r^{-k_{\rho}}$.  Observations of dense cores in Infrared Dark Clouds
find $k_{\rho}\simeq1.3$--$1.6$ \citep{but12,but14}, and we adopt
$k_{\rho}=1.5$ as a fiducial value, which is the same as the fiducial
value used by \citet{mck03} \citep[also][]{zha11,zha13,zha14}.  Then,
the radius of a core is given as
$R_c=0.057(M_{c}/60\:\msun)^{1/2}(\Sigcl/\gcm)^{-1/2}\:{\rm pc}$.  The
core radius is smaller for higher-$\Sigcl$ since the core is in
pressure equilibrium with the ambient clump.  The rotational parameter
is fixed as $\beta_c=0.02$, i.e., similar to values derived from
observations of lower-mass cores \citep{goo93,li12,pal13}.
In this study, we investigate the collapse of
cores with $M_c=10$--$3000\:\msun$ at $\Sigcl=0.1$--$0.316\:\gcm$,
$10$--$1000\:\msun$ at $1\:\gcm$, and $10$--$300\:\msun$ at
$3.16\:\gcm$.

The inside-out collapse of a core that is a singular polytropic sphere
is described by the self-similar solution \citep{mcl97,mck03}, which
gives the infall rate onto the central protostar-disk system in the
limit of no feedback:
\begin{eqnarray}
	\Mdot_{*d}(t)&=&9.2\times10^{-4}
	\left( \frac{M_{*d}(t)}{M_{c}} \right)^{0.5}\nonumber\\
	&\times&
	\left( \frac{M_{\rm c}}{60\msun} \right)^{3/4}
	\left( \frac{\Sigcl}{\gcm} \right)^{3/4}
	\msunyr, \label{eq_Mdot}
\end{eqnarray}
where $M_{*d}(t) =\int \Mdot_{*d} dt$ is the collapsed mass, which
indicates the mass of the protostar and disk if there was no feedback
at all. A higher clump mass surface density leads to a more compact
core and thus a shorter free-fall time and higher infall rate.  Also,
this formula indicates that the infall rate increases with time in the
no-feedback case (set by the choice of $k_\rho=1.5$; a choice of
$k_\rho=2$ would lead to a constant infall rate).
To obtain the actual mass accretion rate, we need to calculate the effect of
feedback processes, which will be described in \S\ref{sec_method_fb}.
We note that, in all models in this study, the accretion rates are
always smaller than the Eddington rate of $\sim2 \times
10^{-2}(r_*/10\rsun)\msunyr$.

Since the initial core is rotating, a disk is assumed to form around
the protostar. For simplicity, we only include the effect of rotation
inside the sonic point where the infall becomes supersonic and assume
that the ratio of the rotational to gravitational energy is constant
at this location, $\beta(<r)=\beta_c$.
Based on the angular momentum conservation from the sonic point,
the disk radius is given by
\begin{eqnarray}
	r_d(t) =156
	\left( \frac{\beta_c}{0.02} \right)
	\left( \frac{M_{*d}(t)}{m_{*d}} \right)
	\left( \frac{M_{*d}}{M_{c}}\right)^{2/3}  \nonumber\\
	\times
	\left( \frac{M_{c}}{60\msun}\right)^{1/2}
	\left( \frac{\Sigcl}{\gcm}\right)^{1/2}
	{\rm AU},\label{eq_rd}
\end{eqnarray}
\citep[see \S 2.1 of][]{zha14}. The protostellar disk is expected to
be massive and self-gravitating due to high mass supply from the
infalling envelope.  The angular momentum is transported efficiently
by torques in such a massive disk \citep[e.g.,][]{per16}, keeping the
mass ratio of disk and protostar approximately constant at
$f_{d}\simeq1/3$ \citep[e.g,][]{kra08}.
We note that these density structures of rotating infall and protostellar disks were developed by
\citet{zha11,zha13,zha14} and were then used in radiative transfer
calculations for synthetic observations. However, in this study, we
only focus on the accretion history of forming massive stars, and thus
do not need the detailed structure of the envelopes and disks, except
for the opening angle of the outflow cavity. 
Thus we do not expect the results to be very sensitive to the choice
of $\beta_c$ as long as $\beta_c\ll1$ so that the outer core structure
is quasi spherical.

The properties of the protstar, such as luminosity, radius, effective
temperature, and their evolution are important to evaluate the
strength of feedback.  In our study, the protostellar evolution is
calculated self-consistently, being adapted to the accretion rate
using the model of \citet{hos09} and \citet{hos10} \citep[which is
  based on the method developed by][]{sta80,pal91}.  Since the typical
mass-accretion rate in massive star formation is higher than that in
low-mass star formation, the rate of entropy carried into the star is
also high.  This leads to a large protostellar radius of
$\sim100\rsun$ before Kelvin-Helmholtz (KH) contraction starts to be
effective \citep{pal91,hos09}. This swelling causes lower effective
temperatures and lower ionizing photon rates than those predicted by
the ZAMS model at the same mass.  The evolution also depends on the
geometry of the accretion flow, i.e., spherical or disk accretion.
The accretion geometry is quasi-spherical when the expected disk
radius $r_{d}$ is smaller than the stellar radius $r_*$.  In this
case, a shock front is produced when this flow hits the stellar
surface and a fraction of the released gravitational energy is
advected into the stellar interior, which is referred as the ``hot''
shock boundary.  On the other hand, if $r_{d}>r_*$, the material
accretes onto the stellar surface through a geometrically thin-disk.
In disk accretion, much of the energy radiates away before the
material settles onto the star. In the limiting case the entropy
carried into star can be assumed to be the same as the gas in the
stellar photosphere, which is referred as the ``cold'' photospheric
boundary condition. In our model, the calculation starts from the hot
shock boundary and switches to the cold photospheric boundary at
$r_{d}=r_*$.

When the accreting material reaches the stellar surface, the accretion
energy of $L_{\rm acc}=G m_*\mdot_{\rm *acc}/(2r_*)$
(in the case of disk accretion) is released,
where $\mdot_{\rm *acc}$\footnote[2]{
The accretion rate on to the star is described as $\mdot_*$
in previous works \citep{zha11,zha13,zha14,tan16}.
However, $\mdot_{\rm *acc}$ is adopted in this study
since the actual mass growth rate of the star is smaller than this 
due to the mass loss by stellar wind,
i.e., $\mdot_*=\mdot_{\rm *acc}-\mdot_{\rm *w}$.}
is the accretion rate onto the star.
Following previous works, we treat this
accretion luminosity and the intrinsic internal stellar luminosity as
radiating isotropically with a single effective temperature: $L_{\rm
  *acc}=L_*+L_{\rm acc}=4\pi r_*^2 \sigma T_{\rm *acc}^4$, where
$\sigma$ is the Stefan-Boltzmann constant.  Following \citet{tan16},
we adopt the stellar atmospheric model ``Atlas'' \citep{cas04} to
obtain the stellar spectrum $L_{\nu, {\rm *acc}}$.  Then the ionizing
photon rate is evaluated as $S_{\rm *acc}=\int^{\infty}_{\nu_{\rm Ly}}
L_{\nu, {\rm *acc}}(h\nu)^{-1}d\nu$.  Due to line absorption, the
ionizing photon rate can be smaller by orders of magnitude than that
simply evaluated by integrating over a blackbody spectrum, especially
when $T_{\rm *acc}\la2\times10^4\:{\rm K}$.

\subsection{Feedback processes}\label{sec_method_fb}

The accretion rate onto the star is smaller than the collapse rate
given by equation (\ref{eq_Mdot}) because of feedback.  It is
necessary to estimate the impact of feedback to obtain the final mass and the SFE.
Here we explain how we evaluate the feedback processes and their effect on the accretion rate.

\subsubsection{Outflow driven by momenta of MHD disk wind and radiation pressure}\label{sec_momentum}

The bipolar outflow sweeps up part of the core and thus helps to set
the SFE.  We calculate the opening angle of the outflow cavity
$\theta_{\rm esc}$ considering momenta of the MHD disk wind and
radiation pressure, i.e., $p_{\rm dw}$\footnote[3]{The subscript ``w''
  was used to represent the MHD disk wind in previous works
  \citep{zha11,zha13,zha14,tan16}.  However, ``dw'' is adopted in this
  study to distinguish with the new component of stellar wind which is
  described by ``*w''.}  and $p_{\rm rad}$. \citet{zha14} included MHD
disk wind feedback using the model of \citet{mat00}. In this model, if
the outflow momentum is strong enough to accelerate the core material
to its escape speed, the outflow extends in that direction.  We simply
extended this model including the additional term of the radiation
pressure: the following equation is satisfied at the polar angle of
$\theta=\theta_{\rm esc}(t)$
\begin{equation}
c_{g}\frac{dM_{c}}{d\Omega}v_{\rm esc}=\frac{d p_{\rm dw}(t)}{d\Omega} + \frac{d p_{\rm rad}(t)}{d\Omega},\label{eq_open}
\end{equation}
where $\Omega$ is the solid angle, $v_{\rm esc}=\sqrt{2GM_c/R_c}$ is
the escape velocity from the core, and $c_{g}$ is a correction factor
to account for the effects of gravity on the propagation of the
shocked shell. Following \cite{zha14}, the angular distribution of the
core mass is assumed to be isotropic: $dM_c/d\Omega=M_c/4\pi$, even
though in reality the core would be expected to flatten to some degree
by rotation and/or large scale magnetic field support.
Based on Appendix of \citet{mat00}, we estimate $c_{g}=2.63$ for our core set up.

The total MHD disk wind momentum $p_{\rm dw}(t)$ is evaluated by
integrating the momentum rate of the wind using a semi-analytic disk
wind solution that is modified from the centrifugally driven MHD
outflow model of \citet{bla82}:
\begin{eqnarray}
\dot{p}_{\rm dw}(t)&=&\phi_{\rm dw}\mdot_{\rm *acc}v_{\rm K*},  \label{eq_pdotdw}  \label{eq_pdw}\\
\phi_{\rm dw}(t)&=&4\sqrt{15}f_{\rm dw}\frac{1-(r_{d}/r_*)^{-1/2}}{\ln(r_{d}/r_*)},
\end{eqnarray}
where
$v_{\rm K*}=\sqrt{Gm_*/r_*}$ is the Keplerian speed at the stellar radius,
$\phi_{\rm dw}$ is the factor to measure the disk wind momentum in terms of $m_{\rm *acc} v_{\rm K*}$ \citep{tan02},
$f_{\rm dw}$ is the mass loading rate of wind relative to the accretion rate onto the star
\citep[see][for derivation]{zha13,zha14}.
We fix the mass loading rate as $f_{\rm dw}=0.1$ as a typical value of disk winds \citep{kon00}.
According to results of our evolution calculation,
we find the typical value of $\phi_{\rm dw}$ is $0.15$--$0.3$.
The angular distribution of the momentum of MHD disk wind is described as
\citep{mat99,shu95,ost97}
\begin{equation}
P(\mu)\tbond \frac{4\pi}{p_{\rm dw}}\frac{d p_{\rm dw}}{d\Omega}=\frac{1}{\ln(2/\theta_0)(1+\theta_0^2-\mu^2)}, \label{eq_phidw}
\end{equation}
where $\theta_0$ is a small angle which is estimated to be $0.01$, and
$\mu=\cos\theta$ (please note that $\int_0^1P d \mu=1$).
This angular distribution of $P(\mu)$ encapsulates the collimated nature of MHD disk winds.
As a result of some trapping by the core, the actual disk wind
mass-loss rate is smaller than $f_{\rm dw}\mdot_{\rm *acc}$,
which is the limiting value for a fully opened cavity.
The fraction of the mass of the wind that can escape from the outflow cavity,
$f_{\rm dw,esc}$, is evaluated based on the fraction of the mass flow
in the directions $0\leq\theta\leq \theta_{\rm esc}$.  \citet{zha14}
derived $f_{\rm dw,esc}$ to be
\begin{eqnarray}
f_{\rm dw,esc}&(&\mu_{\rm esc})=-\frac{2}{\ln(r_{d}/r_*)}\nonumber\\
\times &\ln& \left[\sqrt{\frac{r_*}{r_{d}}} +
\left(1 - \sqrt{\frac{r_*}{r_{d}}} \right) \int_{0}^{\mu_{\rm esc}}  P(\mu) d\mu
\right], \label{eq_dwesc}
\end{eqnarray}
where $\mu_{\rm esc}=\cos\theta_{\rm esc}$.  Then, we have the MHD
disk wind rate as
\begin{eqnarray}
	\mdot_{\rm dw}=f_{\rm dw}f_{\rm dw,esc}\mdot_{\rm *acc}. \label{eq_mdotdw}
\end{eqnarray}
Note that this value is the mass-loading rate from the disk,
however, it is not the total mass-loss by the MHD disk wind from the core.
The momentum by the MHD disk wind (and radiation pressure)
sweeps up much large amount of gas from the envelope creating the outflow cavity.
As we will see in \S \ref{sec_results},
this outflow driven by the disk wind is the most significant feedback.

In low-mass star formation, the MHD disk wind is the dominant source
of momentum feedback. Radiation pressure becomes significant if the
stellar mass reaches $\sim20\msun$. The momentum from radiation
pressure $p_{\rm rad}$ is obtained by the integral of the radiation
pressure momentum injection rate which is given by
\begin{equation}
\dot{p}_{\rm rad}(t)= f_{\rm trap}\frac{L_{* \rm acc}}{c}, \label{eq_prad}
\end{equation}
where $f_{\rm trap}$ is a trapping factor accounting for the increment
of direct radiation pressure force by dust re-emission
\citep{tho05,mur10,mur11}. This radiation by dust re-emission should
be reduced significantly by the pre-existing MHD outflow cavity
\citep{kru05,kui15,kui16} and/or the RT instability \citep{kru09,ros16}.
Therefore, in the implementation of our model in this paper the effect
of dust re-emission is ignored and only direct stellar radiation is
considered, i.e., $f_{\rm trap}=1$. Therefore, since we are only
considering direct stellar radiation, the angular distribution of the
radiation pressure momentum is assumed to be isotropic:
$dp_{\rm rad}/d\Omega=p_{\rm rad}/4\pi$.

Material in the envelope is swept-up by the momenta of MHD wind and radiation pressure as the opening-up of the outflow cavity.
The mass-loss associated with this sweeping process can be evaluated as
\begin{equation}
	\Mdot_{\rm swp}=-\dot{\mu}_{\rm esc}(t)(M_c-M_{*d}(t)),\label{eq_swp}
\end{equation}
where the negative sign is chosen to make the mass-loss rate positive
since $\dot{\mu}_{\rm esc}<0$ (see also \S\ref{sec_acc_rate}).  We
note that it is not straightforward to clearly distinguish the
separate mass-loss contributions here due to MHD disk wind and
radiation pressure since the momentum from these two feedback
mechanisms combine to open-up the outflow cavity. Below, we will
compare the mass-loss from the system by this mechanism with that due
to other feedback processes.

We also include effect of shielding by the inner disk. Since this
inner disk shielding is efficient to overcome the direct stellar
radiation pressure, infall can always continue from the disk shadow
region \citep{tan11,kui12}. Therefore, we limit the maximum opening
angle based on the aspect ratio of inner disk, i.e.,
$\theta_{\rm esc,max}=\tan^{-1}(H/r)$. 
We calculate the inner disk structure with an $\alpha$-disk model
\citep{sha73} using the pseudo-viscosity $\alpha$ parameter, which
depends on self-gravitational stability \citep{TO14}:
$\alpha\rightarrow0.01$ if the disk is stable with respect to
self-gravity (the typical situation) implying turbulence is driven by
the magneto-rotational instability; $\alpha\rightarrow1$ if the disk
is marginally gravitationally unstable implying angular momentum
transport is governed by gravitational torques (but this case does not
arise in our models for the inner disk region of interest).  Note that
angular-momentum transport by the disk wind is not explicitly
considered here in this calculation
\citep[although it was accounted for in the larger scale disk structure calculations
of ][]{zha13,zha14}:
one expects that its effects would be to change the effective
value of $\alpha$. However, the disk scale height is not very
sensitive to $\alpha$, i.e., $H\propto\alpha^{1/10}$, and thus we
consider that our estimate of the angular size of the shielded region
is reasonably well estimated by this method.
The aspect ratio is evaluated at the radius of $r=10r_*$ following \citet{mck08}.
Typically, the disk aspect ratio is about $0.1$ and thus the maximum
opening angle is about $84\degr$.

\subsubsection{Dissipation of envelope and disk by photoevaporation}\label{sec_evp}

The ionizing photon rate $S_{\rm *acc}$ increases dramatically after
KH contraction starts, and the amount of ionizing photons creates an
\HII~region even during the accretion phase. Ionized gas with high gas
pressure can escape from the gravitational binding of the protostellar
core, i.e., photoevaporation.
We have derived a formula of the mass-loss rate by photoevaporation
based on a ray-tracing radiative transfer calculation \citep{tan13},
which is the updated version of the classic analytic model by
\citet{hol94}. However, \citet{tan13} did not consider the effect of
the dust grains since it was applied to the case of primordial star
formation.  Here we extended the photoevaporation model including the
effect of dust attenuation of ionizing photons.

The photoevaporation mass-loss rate $\Mdot_{\rm pe}$ is evaluated
contribution from both the upper and lower surfaces
\citep{hol94,tan13},
\begin{equation}
\Mdot_{\rm pe}= 2\int_{r_g}^{r_0(M_{*d})} 2\pi r X^{-1}m_{\rm H}n_0(r') c_{\rm HII}dr',\label{eq_mdot_pe}
\end{equation}
where $r_g$ is the gravitational radius inside which the ionized gas
is gravitationally bound, $r_0(M_{\rm *d})$ is the collapse radius
inside which the enclosing mass was originally equals to $M_{*d}(t)$,
$n_0(r)$ is the base density at the ionization boundary,
and $c_{\rm HII}$ is the sound speed of the ionized gas.
The gravitational radius in the dust-free
case is determined as the escape velocity becomes comparable to
$c_{\rm HII}$, $r_{\rm g,df} = Gm_{*d}\left(1- \Gamma_{e}
\right)/c_{\rm HII}^2$, where
$\Gamma_{e}=2.6\times10^{-5}(L_{\rm *acc}/\lsun)(m_*/\msun)^{-1}$ is
the Eddington factor for electron scattering \citep{hol94,mck08}.  On
the other hand, the gravitational radius in the dusty case can be
evaluated as the dust-sublimation radius, $r_{\rm sub} =
\sqrt{\kappa_* L_{\rm *acc}/4\pi\sigma\kappa_{\rm sub}T_{\rm sub}^4}$,
where $T_{\rm sub}$ is the dust sublimation temperature which we set
as $1400\:{\rm K}$, and $\kappa_*$ and $\kappa_{\rm sub}$ are the dust
opacity for the stellar radiation and at the dust sublimation
temperature, respectively.  This is because radiation pressure acting
on dust grains assists the ionized gas to become unbounded from the
stellar (and disk) gravity especially when photoevaporation occurs
actively ($\ga20\msun$). Therefore, we evaluate the gravitational
radius as $r_{g}=\min(r_{\rm g,df},r_{\rm sub})$.  The outer boundary
of integration in equation (\ref{eq_mdot_pe}) is chosen as the
collapse radius, considering the evaporation not only from the
protostellar disk but also from the infalling envelope.

The profile of the base density, $n_0(r)$, determines the total
photoevaporation rate (eq. \ref{eq_mdot_pe}).  In the dust-free case,
the radiative transfer calculation by \citet{tan13} showed that the
direct stellar radiation dominates at the ionization boundary, and
derived an analytic formula of $n_0(r)$ in the dust-free case as,
\begin{equation}
n_0(r)= c_{\rm pe}\left(\frac{S_{\rm *acc}}{4\pi\alpha_{\rm A}r^3}\right)^{1/2},
	\ \ {\rm for} \ \  r<r_{\rm sub}, \label{eq_n0_df}
\end{equation}
where $\alpha_{\rm A}$ is the recombination coefficient for all levels
(so-called case A) and $c_{\rm pe}\simeq0.4$ is the correction factor
used to match numerical results. In the dusty region, we extend this
formula including the absorption by dust grains as,
\begin{equation}
n_0(r)= c_{\rm pe}\left(\frac{S_{\rm *acc} e^{-\uptau_{d}}}{4\pi\alpha_{\rm A}r^3}\right)^{1/2},
	\ \ {\rm for} \ \  r>r_{\rm sub}, \label{eq_n0_d}
\end{equation}
where $\uptau_{d}$ is the optical depth caused by dust grains for
ionizing photons evaluated from the dust sublimation radius, i.e.,
\begin{equation}
\uptau_{d}=\int_{r_{\rm sub}}^r n_0(r') \sigma_{\rm a,d} dr', \label{eq_taud}
\end{equation}
and $\sigma_{\rm a,d}$ is the absorption cross sections of dust grains
per H nucleon, which we fix at $10^{-21}\:{\rm cm}^{-2}$ from a
typical value of the diffuse interstellar medium \citep{wei01}
(however, note that the properties of dust in the upper layers of
accretion disks around massive protostars are not well constrained).
As we will see in \S\ref{sec_results}, dust attenuation of ionizing
photons is important for regulating the total photoevaporation rate.
Using this base density profile (eqs. \ref{eq_n0_df} and
\ref{eq_n0_d}), we obtain the photoevaporation rate $\Mdot_{\rm pe}$
integrating the equation (\ref{eq_mdot_pe}).  Please note that, we
calculate the temperature of the ionized gas based on the protostellar
spectrum and the gas density following \citet{tan16}:
it is typically close to $10,000\:{\rm K}$.

We note that our model is not a fully self-consistent unification of
MHD disk wind and photoevaporation feedback, since neither the
magneto-centrifugal acceleration of the photoevaporation flow nor the
photoionization mass-loading of the MHD disk wind are considered.  The
Alfv{\'e}n speed decreases with distance as the Keplererian speed in
the BP wind solution, while the ionized gas sound speed remains
constant at $\sim10\:\kms$.  Therefore, the pure-MHD disk wind should
dominate in the inner region of $r\ll r_{g}$.  On the other hand, in
the outer region where $r\gg r_{g}$, the pure-photoevaporative process
is expected to be most important.  Additionally, gas in the envelope
rotates more slowly than Keplerian, so a magneto-centrifugal wind is
not expected to be efficiently launched from this location.  Thus, our
model is expected to be appropriate at both of the extreme ends of
inner and outer radii. Conventionally, those two flows are discussed
separately.  However, in reality, the mass-loss by thermo- and
magneto-hydrodynamical processes occur together, and a unified model
is necessary for a more accurate treatment, which we defer to a future
paper.  For more discussion about ``magneto-photoevaporation,'' see
\citet{bai16}, who studied the MHD disk wind including far-UV/X-ray
heating in protoplanetary disks.

\subsubsection{Stellar wind mass-loss}\label{sec_stwind}

The mass-loss via a stellar wind driven by radiative forces on
spectral lines is also considered in our model. \citet{vin11} studied
the stellar wind mass-loss rate up to $m_*=300\msun$ based on Monte
Carlo radiative transfer models and dynamically consistent spherical
structure. They found two regimes of stellar wind mass-loss: one is
the normal O-type wind regime for $\Gamma_{e}<0.7$; the other is the
extreme WR wind regime for $\Gamma_{e}>0.7$. The mass-loss rate
dramatically increases with $\Gamma_{e}$ and they called this upturn
at $\Gamma_{e}=0.7$ as the ``kink.''  We adopt a stellar wind
mass-loss rate as a function of stellar mass $m_*$ and luminosity
$L_{*\rm acc}$ based on the fiducial results of \citet{vin11}:
\begin{eqnarray}
\mdot_{\rm *w}&=&6.3\times10^{-7} \left(\frac{m_*}{\msun}\right)^{0.7}
\left(\frac{\Gamma_{e}}{0.7}\right)^{a}\msunyr, \label{eq_mdot_sw} \\
a &=&
\begin{cases}
	2.2 & (\Gamma_{e}<0.7), \\
	4.77 & (\Gamma_{e}>0.7).
\end{cases}
\end{eqnarray}
This mass-loss rate is evaluated based on a fixed effective
temperature of $50,000\:{\rm K}$. \citet{pet16} have suggested that
the mass-loss rate would jump up about one order of magnitude if the
effective temperature is lower than $25,000{\rm K}$. However, our
protostellar evolution calculation shows that the effective
temperature is always higher than $35,000\:{\rm K}$
when the Eddington factor is higher than $0.4$.
Also the variation of mass-loss rate
with effective temperature is less than a factor of a few in this high
temperature range. Thus, the stellar wind mass-loss rate given by
equation (\ref{eq_mdot_sw}) is a reasonable approximation for our
model, even ignoring the $T_{\rm *acc}$ dependence.  Indeed, we will
show that the stellar wind mass-loss has only a minor effect compared
to other feedback processes.

\subsection{Net accretion rate onto stars with feedback}\label{sec_acc_rate}

We have introduced estimations of the impact of multiple feedback
processes.
We now evaluate the accretion rate of stars given the effects of these kinds of feedback.
The total mass of the envelope at a certain moment is $M_{\rm
  env}=\mu_{\rm esc}(t)(M_c-M_{*d}(t))$.  Taking the time-derivative
of $M_{\rm env}$, we get
\begin{eqnarray}
	\Mdot_{\rm env} =
	\dot{\mu}_{\rm esc}\left(M_c-M_{*d}\right) - \mu_{\rm esc}\Mdot_{*d}\nonumber\\
	= -\Mdot_{\rm swp} - \mu_{\rm esc}\Mdot_{*d}.
\end{eqnarray}
The first term on the right hand side is the sweeping rate by the
opening-up of the outflow cavity created by the momenta of the MHD
disk wind and radiation pressure (eq. \ref{eq_swp}). The second term
represents the infall rate onto the star-disk system.
From mass conservation in the infalling flow, we have
\begin{eqnarray}
	\mu_{\rm esc}\Mdot_{d*}=\mdot_*+\mdot_{\rm *w}+\mdot_d+\mdot_{\rm dw}+\Mdot_{\rm pe},
\end{eqnarray}
where $\mdot_d$ is the mass growth rate of the disk (see also
Fig. \ref{fig_schematic}).  Note that, due to the stellar wind
mass-loss, the net accretion rate, or the stellar-mass growth rate, is
smaller than the accretion rate onto the star,
\begin{eqnarray}
	\mdot_* = \mdot_{\rm *acc} - \mdot_{\rm *w}.
\end{eqnarray}
Following our previous study,
the mass ratio of disk and star is assumed to be constant at $f_d=m_d/m_*=1/3$
by the self-gravitational-torque regulation \citep[e.g,][]{kra08},
and thus the disk mass growth rate is $\mdot_{\rm d}=f_d\mdot_{*}$.
Using also equation (\ref{eq_mdotdw}),
the net mass growth rate of the star is
\begin{eqnarray}
	\mdot_* = 
	\frac{\mu_{\rm esc} \Mdot_{*d} - \Mdot_{\rm pe}-(1+f_{\rm dw}f_{\rm dw,esc})\mdot_{\rm *w}}
	{1+f_{d}+f_{\rm dw}f_{\rm dw,esc}}
	. \label{eq_mdot_tot}
\end{eqnarray}
All quantities are time variable except $f_d$ and $f_{\rm dw}$.
Eliminating the terms with $\Mdot_{\rm pe}$ and $\mdot_{\rm *w}$, this
equation is identical to that in \citet{zha14}.
Feedback by radiation pressure does not appear explicitly in equation
(\ref{eq_mdot_tot}), however, it increases the opening angle
$\theta_{\rm esc}$ and escape fraction $f_{\rm dw,esc}$.

We continue the protostellar evolution calculation as long as
$\mdot_{*}(t)>0$, i.e., the stellar mass increases, and determine
the stellar mass at the moment of $\mdot_{*}=0$ as the final mass
when it forms $m_{\rm *f}$. Note that, since the outflow opening angle has
a limit set by the disk shielding effect, the outflow from the MHD
disk wind and radiation pressure cannot stop mass accretion
completely, i.e., $\mu_{\rm esc}>0$. Therefore, the accretion finishes
when (1) mass-loss by photoevaporation and stellar wind is
significant, or (2) the entire core collapses, i.e., $M_{*d}=M_{c}$.
We define the instantaneous SFE as the ratio of
net accretion rate to infall rate without feedback, i.e.,
$\varepsilon_*(t)\equiv\mdot_*(t)/\Mdot_{*d}(t)$, 
and otherwise use ``SFE'' to refer to the ratio of the final stellar
mass when the accretion stops to the initial core mass, i.e., $\sfe\equiv m_{*f}/M_{c}$.
The instantaneous SFE is important since it is in principle observable
for individual protostellar systems.  For example, \citet{zha16}
measured the detailed structure of the HH46/47 molecular outflow using
Atacama Large Millimeter/sub-millimeter Array (ALMA), and reported the
instantaneous SFE to be $1/4$--$1/3$.  However here, we focus mainly
on the final SFE rather than the instantaneous SFE to discuss the
relation between the CMF and IMF (\S \ref{sec_impact}).

Note that the disk mass is so far ignored in the evaluation of the
final stellar mass.  However, some amount of disk accretion would
still be able to continue even after the entire core collapses.
The accretion rate is expected to decline as the disk mass to stellar
mass ratio drops and self-gravitational torques become less effective.
We expect that such a lower accretion rate disk would be more readily
dissipated by photoevaporation and/or radiation pressure.  However,
the fraction of the disk mass that finally accretes onto the star is
uncertain because the actual angular momentum transport processes are
uncertain at this stage.  Therefore, for simplicity, we ignore the
disk mass in the SFE evaluation, and note that the actual SFE may be
underestimated by up to a factor of $1+f_d\rightarrow4/3$.

\subsection{Primordial star formation}\label{sec_popIII_model}

Although the main purpose of this paper is the study of feedback in
massive star formation in present-day universe, we also apply the same
feedback model to primordial star formation in the early universe for
comparison and demonstration of our model. Here we describe
modifications of the present-day massive star formation model for its
application to primordial star formation. These modifications follow
the methods of \citet{tan04}, \citet{tan04b} and \citet{mck08} for
primordial star formation.

\citet{tan04} predicted the evolution of the mass infall rate,
accretion disk structure, and protostellar evolution associated with
primordial star formation.  We use the results of \citet{tan04} for
the infall rate and disk evolution. The infall rate excluding effects
of feedback is given by
\begin{eqnarray}
	\Mdot_{*d}(t)=0.026 K'^{15/7}\left(\frac{M_{*d}(t)}{\msun} \right)^{-3/7}\msunyr.
	\label{eq_Mdot_popIII}
\end{eqnarray}
Here $K'$ is the entropy parameter of the polytropic equation of state
of the cloud; larger values of $K'$ correspond to denser gas cores.
In this study, the entropy parameter is fixed at the fiducial value of
$K'=1$. 

The above infalling rate replaces that of equation (\ref{eq_Mdot})
that is used to model present-day star formation.
It is interesting that the typical infall rates in primordial star formation and
present-day massive star formation are coincidentally of the same
order. The high accretion rate in the primordial case is induced by
the high gas temperature in the core due to inefficient cooling at
zero metallicity, while for the present-day case the high turbulence
and strong magnetic fields enhance the effective pressure of cores
leading to their high accretion rates. 

Based on the conservation of angular momentum from the sonic point to
the outer radius of disk, the disk radius around Pop III protostars is
evaluated as
\begin{equation}
r_{d}(t) = 3.44 K'^{-10/7}\left(\frac{f_{\rm Kep}}{0.5}\right)^2\left(
\frac{M_{*d}(t)}{\msun}\right)^{9/7}\:{\rm AU},\label{eq_rd_popIII}
\end{equation}
where $f_{\rm Kep}$ is a angular momentum parameter of infalling gas,
which is fixed at the fiducial value of $0.5$ \citep{abe02,tan04}.

As in the case of present-day massive star formation, protostellar
evolution is calculated self-consistently adapted to the accretion
rate using the code developed by \citet{hos09} and \citet{hos10},
except the opacity is modified for zero metallicity. As a result of
similar accretion rates of $10^{-3}\:\msunyr$, the evolution of
primordial protostars is expected to resemble that of present-day
massive protostars \citep{omu01,omu03}. The main difference of
protostellar evolution is that stellar radius in the main-sequence
phase is smaller at zero metallicity than that at solar metallicity.
This is because, due to the lack of C and N, the KH contraction
continues until the temperature becomes high enough for small amounts
of carbon to be produced by He burning, which then enables the
operation of the CNO cycle. For more details of comparison of
protostellar evolution in zero and solar-metallicities, see \S3.4 of
\citet{hos09}.

For our modeling, we also update the stellar spectrum appropriate for
the case of zero metallicity \citep{sch02} in order to evaluate the
ionizing photon rate.  This leads to higher ionizing photon
luminosities for a given temperature due to a lack of metal line
absorption.

The feedback model also needs some modifications to apply to
primordial star formation. In the outflow feedback, we neglect the
momentum by radiation pressure, i.e., $p_{\rm rad}=0$ in equation
(\ref{eq_open}), since there are no dust grains. Following
\citet{tan04b}, we do include the MHD disk wind momentum, since the
MHD disk wind could be driven by the disk-dynamo generated magnetic
field. Due to the different density profile inside the core, the
correction factor accounting for effects of gravity on shock
propagation is $c_{g}\simeq4.6$. The escape velocity is evaluated by
$v_{\rm esc,c}=3.22K'^{5/7}(M_{*d}/1000\msun)^{-1/7}{\rm km~s^{-1}}$
\citep{tan04b}. In the photoevaporation feedback calculation, dust
attenuation is set to zero, $\uptau_{d}=0$. This means
photoevaporation is more efficient in primordial star formation.
Finally, we neglect the the stellar wind mass-loss (i.e.,
$\mdot_{\rm *w}=0$), which is mainly driven by the metal lines.

%%%%%%%%%% Section 3 %%%%%%%%%%
\section{Results} \label{sec_results}

In this section we first present the general evolution of massive
formation by core accretion. Then, we examine details of individual
feedback process. Next, we show the results of primordial star
formation to demonstrate the effect of metallicity.  Finally, we show
the obtained SFE for various initial cores.

\subsection{Accretion history and protostellar evolution}\label{sec_evo}

\begin{figure*}
\begin{center}
\includegraphics[width=160mm]{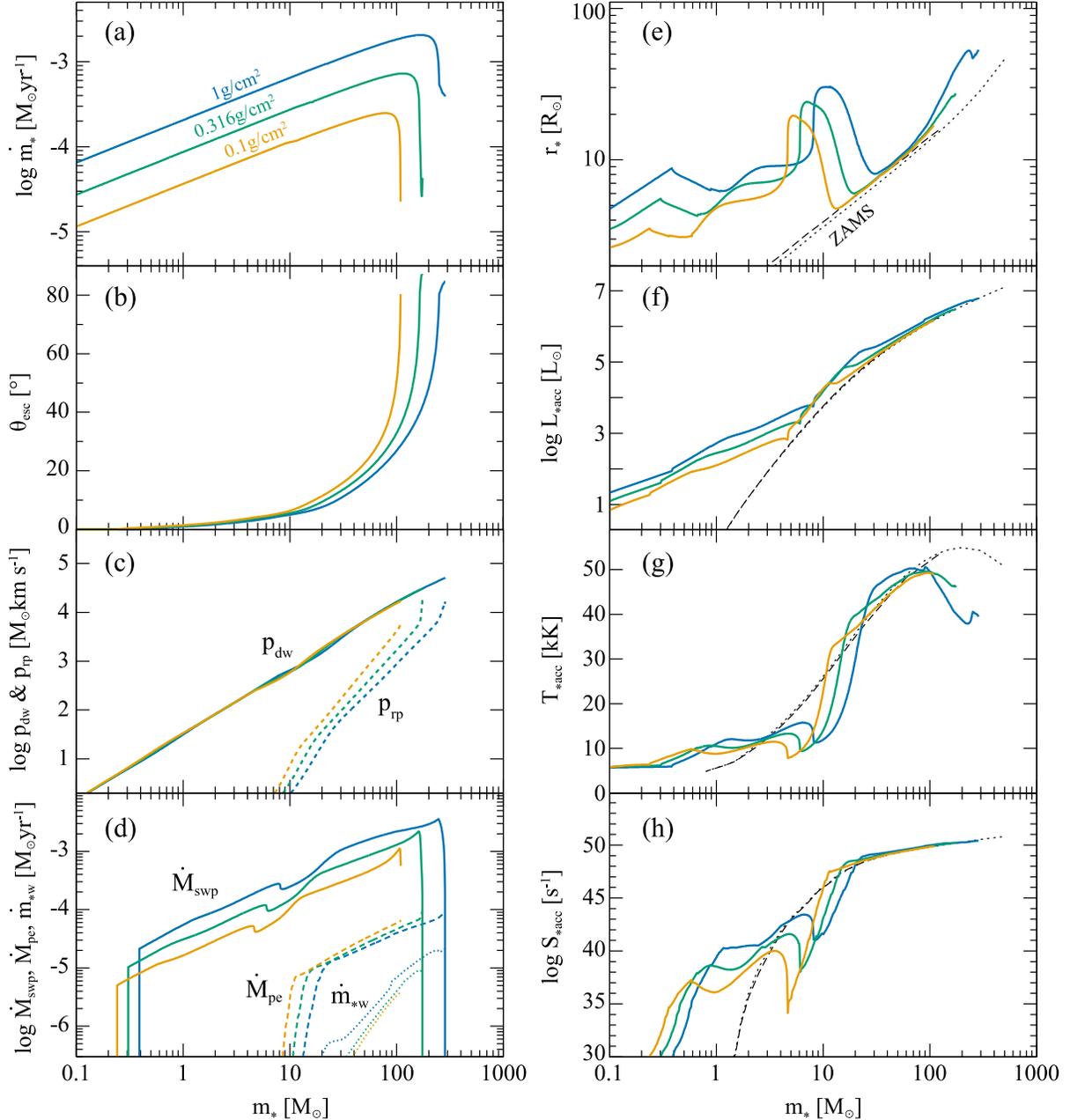}
\end{center}
\caption{
Evolution of protostellar and feedback properties as functions of the
protostellar mass $m_*$ for initial core mass of $M_{c}=1000\msun$ and
$\Sigcl=0.1,~0.316,~{\rm and}~1~\gcm$ (orange, green and blue lines,
respectively).
The panels show: (a) net accretion rate, $\mdot_{*}$; (b) outflow cavity
opening angle, $\theta_{\rm esc}$; (c) momenta of MHD disk wind,
$p_{\rm dw}$ (solid), and radiation pressure, $p_{\rm rad}$ (dashed);
(d) outflow sweeping rate, $\Mdot_{\rm swp}$ (solid),
photoevaporation mass-loss rate, $\dot{M}_{\rm pe}$ (dashed), and
stellar wind mass-loss rate, $\mdot_{\rm sw}$ (dotted); (e) stellar
radius, $r_*$; (f) effective temperature, $T_{*\rm acc}$; (g) total
luminosity, $L_{*\rm acc}$; (h) ionizing photon rate, $S_{\rm *acc}$.
In the right-hand panels, the ZAMS properties are also plotted by
black dash \citep{sch92} and dotted \citep{eks12} lines.}
\label{fig_hist}
\end{figure*}

Figure \ref{fig_hist} shows results of our modeling of present-day
protostars forming from cores with initial mass $M_{c} =
1000\:\msun$. Three different clump mass surface densities are
considered.

First, consider the case with $\Sigcl=1~\gcm$.  As the infall rate
increases (Fig.~\ref{fig_hist}a), the net accretion rate also
increases to $2\times10^{-3}\:\msunyr$ until the stellar mass reaches
$170\:\msun$ (at $t=1.35\times10^5\:{\rm yr}$). Then the accretion
rate drops, finally stopping at $285\:\msun$ ($t=2.64\times10^5\:{\rm
  yr}$).  The decline of accretion is mainly caused by the opening-up
of the outflow cavity (Fig.~\ref{fig_hist}b) given the increasing
momentum of the disk wind and from radiation pressure
(Fig.~\ref{fig_hist}c), rather than by mass-loss by photoevaporation
or via the stellar wind (Fig.~\ref{fig_hist}d).
It is clearly seen that the outflow sweeping rate is orders of
magnitude higher than other mass-loss rates in Figure~\ref{fig_hist}d,
which indicates the MHD outflow, assisted by radiation pressure, is
the most dominant feedback process.
The MHD disk wind always
dominates total momentum, however, the radiation pressure also can
give significant assistance to open-up the outflow cavity (see
\S\ref{sec_outflow} for more details). Mass-loss by photoevaporation
quickly rises to $\sim10^{-5}\:\msunyr$ when the stellar mass is about
$15\msun$, and then increases to $\sim10^{-4}\:\msun$ gradually after
that.  However, the photoevaporation mass-loss rate reaches a maximum
of about $10^{-4}\:\msunyr$, which is never enough to shut down
accretion.  Mass-loss by the stellar wind is even smaller than that
from photoevaporation (thus, the accretion rate onto the star is
almost equal to the net accretion rate, i.e., $\mdot_{\rm
  *acc}=\mdot_*$).  Therefore, mass accretion only finishes when the
entire initial core collapses.  The SFE in this case is
$\sfe=285\msun/1000\msun=0.285$.  Please remind that our evaluation of
the final mass ignores the disk mass, and this SFE is the minimum
estimation with the maximum error of $\Delta \sfe=0.095$ (\S
\ref{sec_acc_rate}).

The evolution of protostellar properties are shown in the right panels
of Figure \ref{fig_hist}. At around $m_*=8\msun$, the stellar radius
(Fig.~\ref{fig_hist}e) suddenly increases by a factor of three which
is due to the redistribution of entropy in the protostar
\citep{hos09,hos10}. The protostar reaches the local maximum radius of
$30\:\rsun$ at $m_*=11\:\msun$.  Until this time the total luminosity
(Fig.~\ref{fig_hist}f) is dominated by accretion luminosity, i.e., the
total lumininosity is significantly larger than the ZAMS
luminosity. The effective temperature of the protostar
(Fig.~\ref{fig_hist}g) is relatively low due to this large stellar
radius. Therefore the ionizing photon rate (Fig.~\ref{fig_hist}h),
which is very sensitive to the effective temperature, is lower than
that of ZAMS model by about five orders of magnitude. 

At later times and greater masses the protostar undergoes KH
contraction and approaches the main sequence structure.  The effective
temperature increases to $45,000\:{\rm K}$, and thus the ionizing
photon rate also dramatically rises leading to the start of
significant photoevaporation.  The star evolves almost along the ZAMS
line at the mass range of $30$--$100\:\msun$. Then, the stellar radius
again becomes slightly larger than the ZAMS model.  This deviation is
related to the metal opacity near the stellar surface and the high
accretion rate of $10^{-3}\:\msunyr$ \citep[see][]{ish99,gra12}.  We note
that this small deviation from the ZAMS has little effect on the
significance of feedback and on the final stellar mass.

Next, we consider the cases of 1000$\:M_\odot$ cores in lower mass
surface density environments, i.e., protostars with lower accretion rates
(see green and orange lines in Fig. \ref{fig_hist}).
We find that the impact of radiation feedback becomes more
significant.
The momentum input due radiation pressure at a given stellar mass is
higher in the lower $\Sigcl$ cases, while the MHD disk wind momentum
is almost identical for all cases.  Due to this higher radiation
pressure momentum, the outflow opens up at lower masses in these lower
$\Sigcl$ cases. Photoevaporation also has a larger impact since the
accretion rate is lower. Especially in the case of $\Sigcl=0.1\:\gcm$,
photoevaporation shuts down mass accretion before the entire core
collapses, i.e., $M_{*d}<M_{c}$. In this way, the relative importance
of radiative feedback becomes higher and results in lower SFE, i.e.,
$\sfe=0.29,~0.18$ and $0.087$ for $\Sigcl=1,~0.316$ and $0.1~\gcm$,
respectively.

The phases of protostellar evolution are shifted to lower stellar
masses for the lower $\Sigcl$ cases because of their lower accretion
rates, which thus mean it takes a longer time for the protostars to
reach a given mass.  However, after the KH contraction phase, the
protostellar evolution following the ZAMS structure very closely in
all cases, at least up to $\sim100\:M_\odot$.

\subsection{Individual feedback processes}

Here we describe results concerning each feedback process included in
the modeling.

\subsubsection{MHD disk wind and radiation pressure driven outflows}\label{sec_outflow}

The outflow in our model is driven by the momenta of the MHD disk wind and by radiation pressure.
As we have seen, the total momentum is dominated by the MHD disk wind.
However, radiation pressure also plays a role in helping to open up the outflow cavities,
since it acts more isotropically than the collimated disk wind.

\begin{figure}
\begin{center}
\includegraphics[width=85mm]{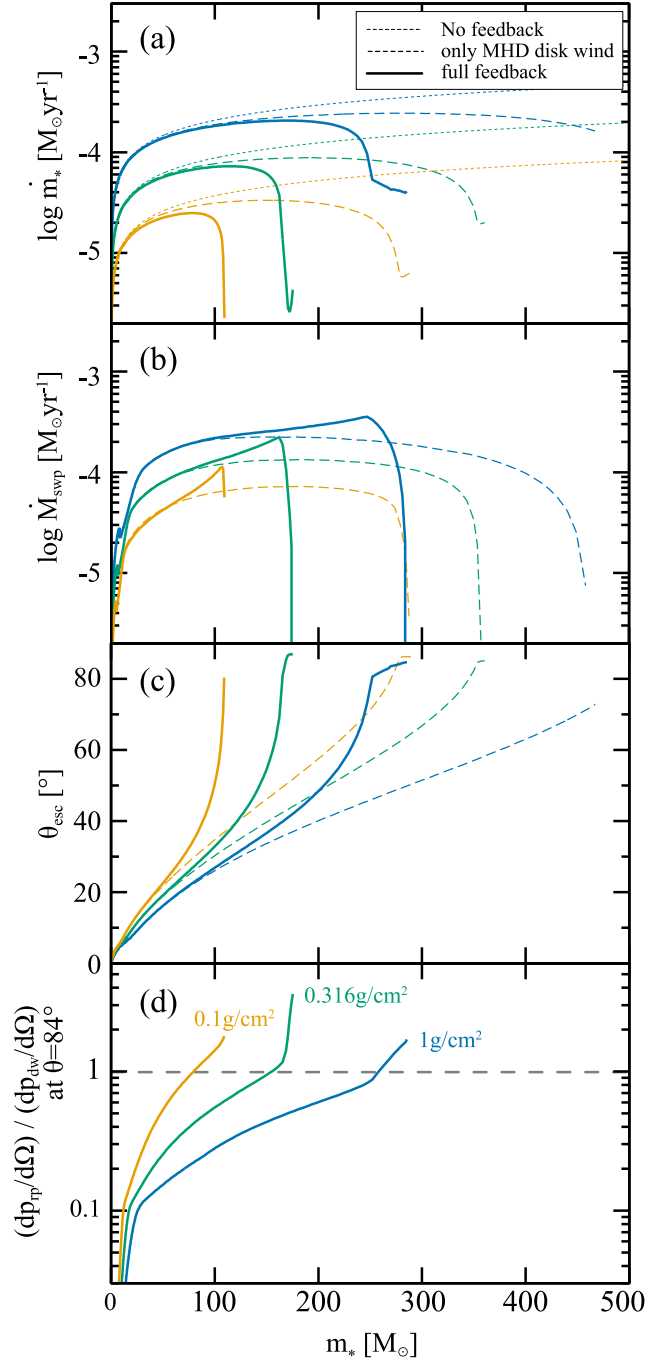}
\end{center}
\caption{
(a) Accretion rate history as a function of protostellar mass
for full feedback models with $M_{c}=1000\msun$ and $\Sigcl=0.1$, $0.316$ and $1\:\gcm$
(orange, green and blue solid lines,
respectively).  The results of these cases with no feedback (dotted)
and those with only MHD disk wind feedback (dashed) are also shown.
(b) Evolution of the mass-loss rate from the envelope due to outflow sweeping,
i.e., opening-up of the outflow cavity, for the same full feedback and only MHD disk wind feedback models.
(c) Evolution of the outflow opening angle $\theta_{\rm esc}$ for the same
full feedback and only MHD disk wind feedback models.  (d) Evolution
of the ratio of momenta from radiation pressure and from the MHD disk
wind at $\theta_{\rm esc}=84\degr$, i.e., $\left(dp_{\rm
  rp}/\ d\Omega\right)/\left(dp_{\rm
  dw}/\ d\Omega\right)|_{\theta_{\rm esc}=84\degr}$, for the same full
feedback models shown above.}
\label{fig_mdot}
\end{figure}

Figure~\ref{fig_mdot}a shows the evolution of
net accretion rates as the protostars grow in mass for models with
$M_{c}=1000\msun$ and $\Sigcl=0.1$, $0.316$, and $1~\gcm$.  For
comparison, the results with only MHD disk wind feedback and those
with no feedback are also shown.  For the case of $\Sigcl=1~\gcm$ with
no feedback, the SFE is unity and thus the stellar mass can reach the
core mass of $1000\:\msun$.  Including MHD disk wind feedback,
the accretion rate is lowered due to the outflow and accretion stops at
$m_*=470\:\msun$.
In the case of all feedback, accretion drops significantly after
$m_*\simeq200\:\msun$ and is finished by $m_*=285\msun$.
The plateau of the accretion rate around $250\:\msun$
is caused by disk shielding.  The SFE including all feedback ($\sfe=0.285$)
is reduced compared to that resulting with only MHD disk wind feedback ($\sfe=0.470$).
As described above, the mass loss by photoevaporation and stellar wind is not very significant,
and the decline of SFE is mainly due to the radiation pressure and its effect on opening up the outflow cavity.
Figure~\ref{fig_mdot}b shows the evolution of the mass-loss rates from
the envelope due to outflow sweeping, i.e., opening-up of the outflow
cavity. When $m_*\la100\msun$, the full feedback models are similar to
the only MHD disk wind models.  However, in the higher mass regime,
the sweeping rate in the full feedback model becomes higher than that
in the only MHD disk wind model and then quickly drops off.  This
phenomenon shows that the outflow cavity opening rate is enhanced by
radiation pressure and more quickly reaches the limit set by disk
shielding.  This can be also seen in Figure~\ref{fig_mdot}c: the
cavity opening in the full feedback models accelerates when
$\theta_{\rm esc}$ reaches about $30\degr$.

The MHD disk wind dominates the total momentum, however, its angular
distribution is highly collimated near the outflow axis
(eq. \ref{eq_phidw}).  On the other hand, the radiation pressure is
modeled as having an isotropic momentum distribution. Therefore, while
the MHD disk wind initially creates the outflow cavity, radiation
pressure has a significant impact in making it wider.
Figure \ref{fig_mdot}d shows the ratio of momenta due to radiation
pressure and the MHD disk wind at an angle of $\theta=84\degr$ (close
to the maximum angle allowed given disk shielding). In the case of
$\Sigcl=1\:\gcm$, the contribution of radiation pressure becomes
similar to that of the disk wind at around $200\:\msun$, which causes
the accretion rate to start falling.
This reduction of accretion rate also
leads to the decline of the momentum rate by the disk wind
since it is accretion powered (eq. \ref{eq_pdotdw}), and the relative
importance of radiation pressure increases even more. In the cases of
lower $\Sigcl$, the radiation pressure has a larger impact, starting
to dominate at lower stellar masses. This is because the lower
$\Sigcl$ leads to lower accretion rates and lower momentum injection
rates from the disk wind, while the radiation pressure does not
strongly depends on the accretion rate. In this way, radiation
pressure has an important impact on the decline of SFE even though the
MHD disk wind dominates the total outflow momentum.

\subsubsection{Mass loss by photoevaporation} \label{sec_pe}

In the models shown in \S\ref{sec_evo}, photoevaporation is not a
significant feedback in setting the SFE, even when $m_*>100\:\msun$
since the photoevaporation mass-loss rate is only
$\sim10^{-4}\:\msunyr$ at its maximum, while the accretion rate is
$\ga10^{-3}\:\msunyr$.  Here we show the importance of dust
attenuation of ionizing photons in the reduction of photoevaporation
feedback.

\begin{figure}
\begin{center}
\includegraphics[width=85mm]{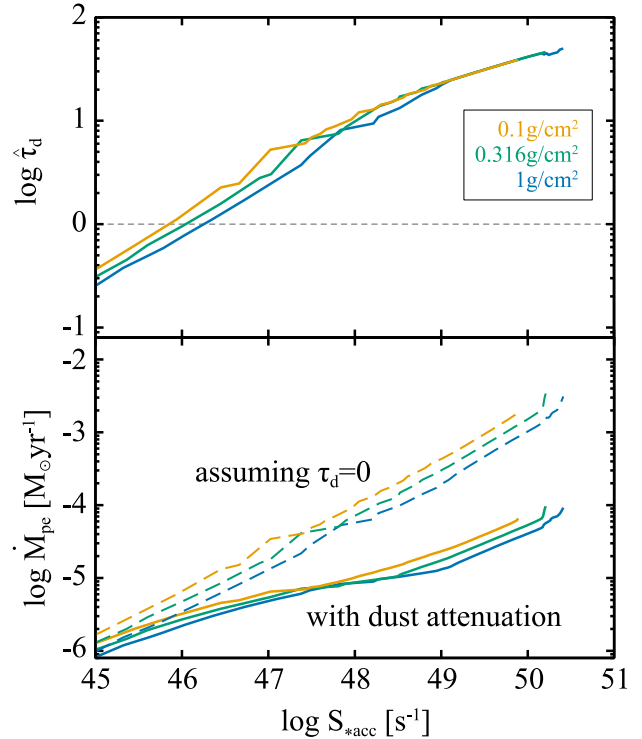}
\end{center}
\caption{
The characteristic optical depth $\hat{\uptau}_{d}$ (top) and
photoevaporation mass-loss rate (bottom) as functions of ionizing
photon production rate for models with $M_{c}=1000\:\msun$ and
$\Sigcl=0.1$, $0.316$ and $1\:\gcm$ (orange, green and blue lines,
respectively). In the bottom panel, hypothetical photoevaporation
mass-loss rates that are evaluated ignoring dust attenuation are also
plotted (dashed lines).}
\label{fig_pe}
\end{figure}

The calculation of the photoevaporation mass-loss rate in our model
includes the effect of dust attenuation on the propagation of ionizing
photons using the optical depth $\uptau_{d}(r)$ (\S\ref{sec_evp}). To
measure the effect of dust attenuation, we introduce a characteristic
optical depth of the system as $\hat{\uptau}_{d}= \sigma_{\rm
  a,d}n_0(r_{\rm sub})r_{\rm sub}$.  Note that, as we will see in
\S\ref{sec_relative}, this characteristic optical depth is not exactly
the same as the total optical depth of the flow $\uptau_{d}(\infty)$,
however it gives a good indication of the optically thin/thick
boundary and well represents the effect of dust attenuation. In
Figure~\ref{fig_pe} we show the characteristic optical depth
$\hat{\uptau}_{d}$ and the photoevaporation mass-loss rate as
functions of ionizing photon production rate for models with
$M_{c}=1000\:\msun$. It can be seen that the characteristic optical
depth increases with $S_{\rm *acc}$ and $\Mdot_{\rm pe}$, and reaches
the optically thick regime when $\Mdot_{\rm
  pe}\simeq2\times10^{-6}\:\msunyr$, which is much smaller than the
typical infall rate.  Even in the optically thick regime, the
photoevaporation mass-loss rate still increases with $S_{\rm *acc}$,
however it does not reach $\sim10^{-3}\:\msunyr$, which is needed to
be a significant feedback effect.

In the bottom panel of Figure~\ref{fig_pe}, to illustrate the
importance of dust attenuation on the photoevaporation mass-loss rate,
we also plot hypothetical rates $\Mdot_{{\rm pe},\uptau_{d}=0}$, which
are evaluated neglecting dust attenuation, i.e., assuming
$\uptau_{d}(r)=0$. In the optically thin regime with
$\hat{\uptau}_{d}<1$, the rates with and without dust attenuation are
similar. However, in the optically thick regime with
$\hat{\uptau}_{d}>1$, the actual photoevaporation mass-loss rate
becomes much smaller than $\Mdot_{{\rm pe},\uptau_{d}=0}$. As we
discuss in \S\ref{sec_relative}, we find that the reduction of
photoevaporation mass-loss rate by dust attenuation can be
approximately described as $\Mdot_{\rm pe}/\Mdot_{{\rm
    pe},\uptau_{d}=0}\simeq1/\hat{\uptau}_{d}$ in the case of
$\hat{\uptau}_{d}\gg1$.  The reduction factor becomes more than one
order of magnitude at high ionizing photon rates of
$\gtrsim10^{49}\:{\rm s^{-1}}$, when the hypothetical mass-loss rate
without dust attenuation would exceed a few$\times10^{-4}\:\msunyr$.
Thus, dust attenuation is very important in limiting the impact of
photoevaporation feedback in present-day massive star formation.

\subsubsection{Mass loss by stellar winds}

\begin{figure}
\begin{center}
\includegraphics[width=85mm]{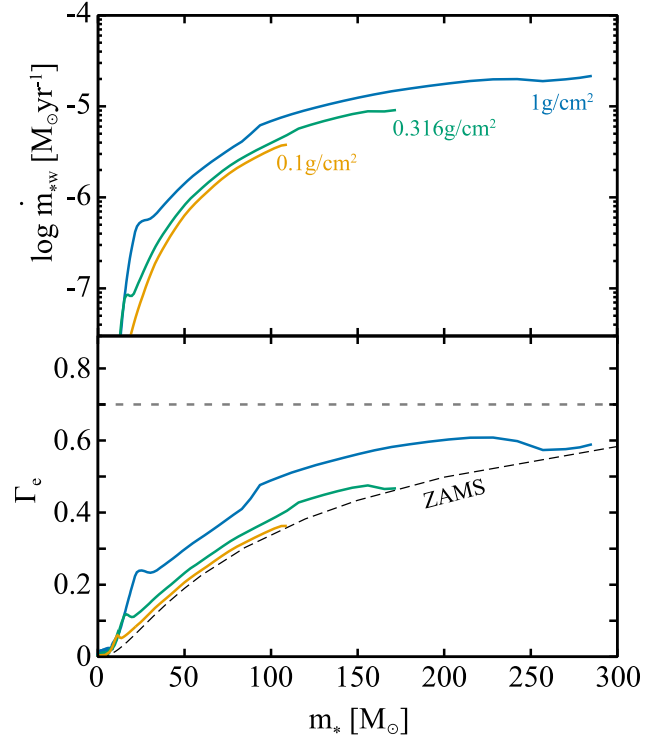}
\end{center}
\caption{
The stellar wind mass-loss rate (top) and the Eddington factor,
$\Gamma_{e}$ (bottom), as functions of protostellar mass, $m_*$, for
models with $M_{c}=1000\:\msun$ and $\Sigcl=0.1$, $0.316$ and
$1\:\gcm$ (orange, green and blue lines, respectively). In the bottom
panel, the Eddington factor evaluated for the ZAMS model \citep{eks12}
is shown by the dotted line, and the critical value of 0.7 is
indicated by the horizontal dashed line.}
\label{fig_sw}
\end{figure}

Mass-loss via stellar winds is a minor effect compared with other
processes. Figure~\ref{fig_sw} shows the stellar wind mass-loss rate
and the Eddington factor $\Gamma_{e}$ as functions of protostellar
mass from results of models with $M_{c}=1000\:\msun$.  The stellar
wind mass-loss rate is about $10^{-5}\:\msunyr$ even at
$m_*=250\:\msun$, which is much smaller than typical values of
accretion rate and photoevaporation rate.  The obtained Eddington
factor with protostellar evolution calculation is slightly higher than
that from the ZAMS model.  However, it is not high enough to reach the
critical ``kink'' value of $\Gamma_{e}=0.7$, above which the mass-loss
rate dramatically increases \citep{vin11}.  Therefore, we conclude
that mass-loss by stellar winds is not a significant feedback effect for setting the SFE.

Note that the momentum input from the stellar wind has been ignored in
our estimation of the outflow opening angle. The stellar wind momentum
rate can be evaluated as $\dot{p}_{\rm sw}\simeq \mdot_{\rm sw} v_{\rm
  *esc}$, where $v_{\rm *esc}=\sqrt{2Gm_*(1-\Gamma_{e})/r_*}$ is the
escape velocity from the stellar surface.  We find that the stellar
wind momentum is at most $10\%$ of the radiation pressure component,
and no more than about 1\% of the MHD disk wind component. We thus
expect that the stellar wind would anyway be confined and collimated
by the MHD disk wind, so would not significantly impact the opening of
the outflow cavity.

\subsection{Primordial star formation}\label{sec_popIII_results}

\begin{figure*}
\begin{center}
\includegraphics[width=170mm]{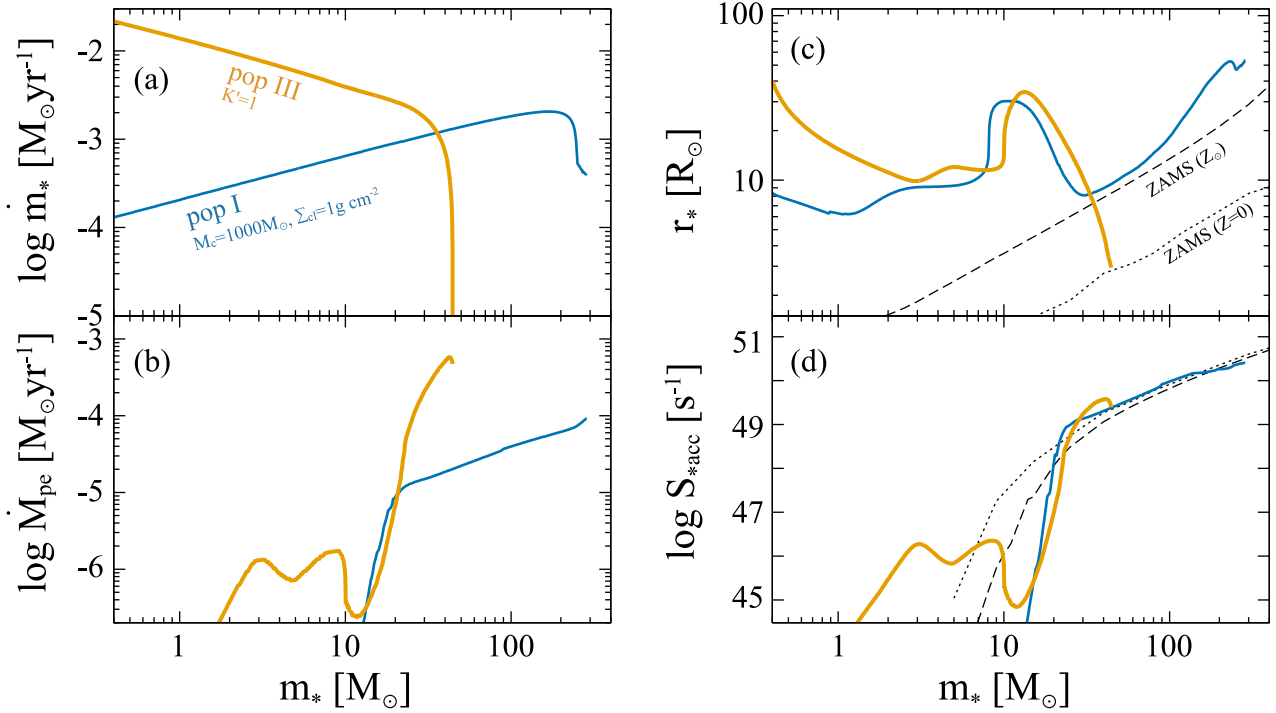}
\end{center}
\caption{
Comparison of primordial star formation ($K'=1$, orange lines) and
present-day massive star formation ($M_c=1000\:\msun$ and
$\Sigcl=1\:\gcm$, blue lines): (a) net accretion rates, $\mdot_{*}$; (b)
photoevaporation mass-loss rates, $\dot{M}_{\rm pe}$; (c) stellar
radii, $r_{*}$; (d) ionizing photon production rates, $S_{\rm
  *acc}$. In panels c and d, the ZAMS models are also shown for
reference: black dashed lines for zero metallicity \citep{sch02} and
black dotted lines for solar metallicity \citep{eks12}.}
\label{fig_zero}
\end{figure*}

We apply our model also for the case of primordial star formation,
which gives a limiting case for the effects of metallicity on massive
star formation feedback
(\S\ref{sec_popIII_model}). Figure~\ref{fig_zero} shows the comparison
of primordial star formation ($K'=1$) and present-day massive star
formation ($M_{c}=1000\:\msun$ and $\Sigcl=1\:\gcm$).  Due to
differences of core density structure, the evolution of accretion rate
is different: the accretion rate decreases with time in primordial
star formation, while it increases in present-day massive star
formation (see Fig \ref{fig_zero}a and eqs. \ref{eq_Mdot} and
\ref{eq_Mdot_popIII}). Therefore, differences between the cases are
not only due to metallicity. However, since the accretion rates are in
fact quite similar at $\sim10^{-3}\msunyr$ when the protostar has
$\sim30\:M_\odot$ an approximate comparison to see the effects of
metallicity is possible.

We find that accretion stops at $44.4\:\msun$ in the primordial case
with $K'=1$, which is a much lower mass than we found for the
present-day case with $M_{c}=1000\:\msun$ and $\Sigcl=1\:\gcm$. The
main reason for this is the high photoevaporation mass-loss rate from
the primordial protostar (Fig. \ref{fig_zero}b). As described in
\S\ref{sec_pe}, dust attenuation of ionizing photon strongly regulates
the photoevaporation mass-loss rate to be $\lesssim10^{-4}\msunyr$ at
solar metallicity.  However, at zero metallicity the photoevaporation
mass-loss rate can reach $\sim10^{-3}\:\msunyr$.  

One may suppose that this difference of photoevaporation mass-loss
rate in the zero and solar metallicity cases is also related to
differences in protostellar evolution and the stellar spectra. As
described in \S\ref{sec_popIII_model}, the primordial protostar
contracts to a smaller ZAMS structure than the solar metallicity
protostar because of its initial lack of heavy elements. The evolution
of stellar radii is shown in Figure~\ref{fig_zero}c, showing that the
primordial protostar contracts to a smaller size after
$m_*\sim30\:M_\odot$. This smaller radius causes higher effective
temperatures and thus higher ionizing photon production rates (also
aided by the lack of metal line absorption in the stellar atmosphere).
However, these differences do not dramatically increase the ionizing
photon rate at zero metallicity (Fig.~\ref{fig_zero}d). When
$m_*\la10\:\msun$, the ionizing photon rate is higher at zero
metallicity than that at solar metallicity, because of higher
accretion rates and luminosities in this earlier phase
(Fig. \ref{fig_zero}a). At the higher mass range of $m_*\ga20\:\msun$,
the difference of ionizing photon rates by metallicity becomes modest
(indeed, the solar metallicity case has even slightly higher ionizing
photon production rates at $15$--$30\:\msun$ due to its smaller radius
during this phase). The ionizing photon production rate difference is
less than a factor of three at $m_*\sim40\:\msun$, which is not enough
to explain the one order of magnitude difference in photoevaporation
mass-loss rate (see bottom panel of Fig. \ref{fig_pe} and
eq. \ref{eq_n0_d}).
Therefore, we conclude that dust attenuation of ionizing photons is
the most significant effect controlling the metallicity dependence of
photoevaporation mass-loss rates.

Our model finds a smaller final stellar mass of $44\:M_\odot$ than the
study of \citet{mck08}, who found $140\msun$ for the $K'=1$
case. The main difference is that we have included MHD disk wind
feedback. As we have seen in the case of present-day massive star
formation, the MHD disk wind is the dominant feedback in the low-mass
regime. This reduction of accretion rate at lower masses results in an
earlier start of KH contraction and thus of effective photoevaporation
feedback. Another difference also results from our updated
protostellar evolution calculation and photoevaporation model compared
with that of \citet{mck08}. For protostellar evolution, we use a
detailed protostellar structure calculation code, which tends to
predict a smaller protostellar size: e.g., at $m_*=30\:M_\odot$ the
model of \citet{mck08} had $r_*\sim20\:R_\odot$
\citep[see Fig.2 of][]{tan08}
while we now estimate $r_*\simeq10\:R_\odot$.
For the photoevaporation calculation, we adopt the model by
\citet{tan13}, who showed the importance of photoevaporation from the
outer region based on an accurate radiative transfer calculation,
while the analytic \citet{hol94} model suggested the mass-loss rate is
dominated by the region close to the inner gravitational radius
$r_{g}$. Including mass-loss from the outer region, including also the
collapsing envelope that is exposed by the outflow cavity, the
photoevaporation rate can be higher by a factor of
$(R_{c}/r_{g})^{0.5}\sim10$ than the Hollenbach et
al. model \citep{tan13}. These differences lead to an enhancement of
feedback compared to the study of \citet{mck08} and result in a
smaller final mass than found in their model.

\subsection{Star formation efficiency}

\begin{figure*}
\begin{center}
\includegraphics[width=180mm]{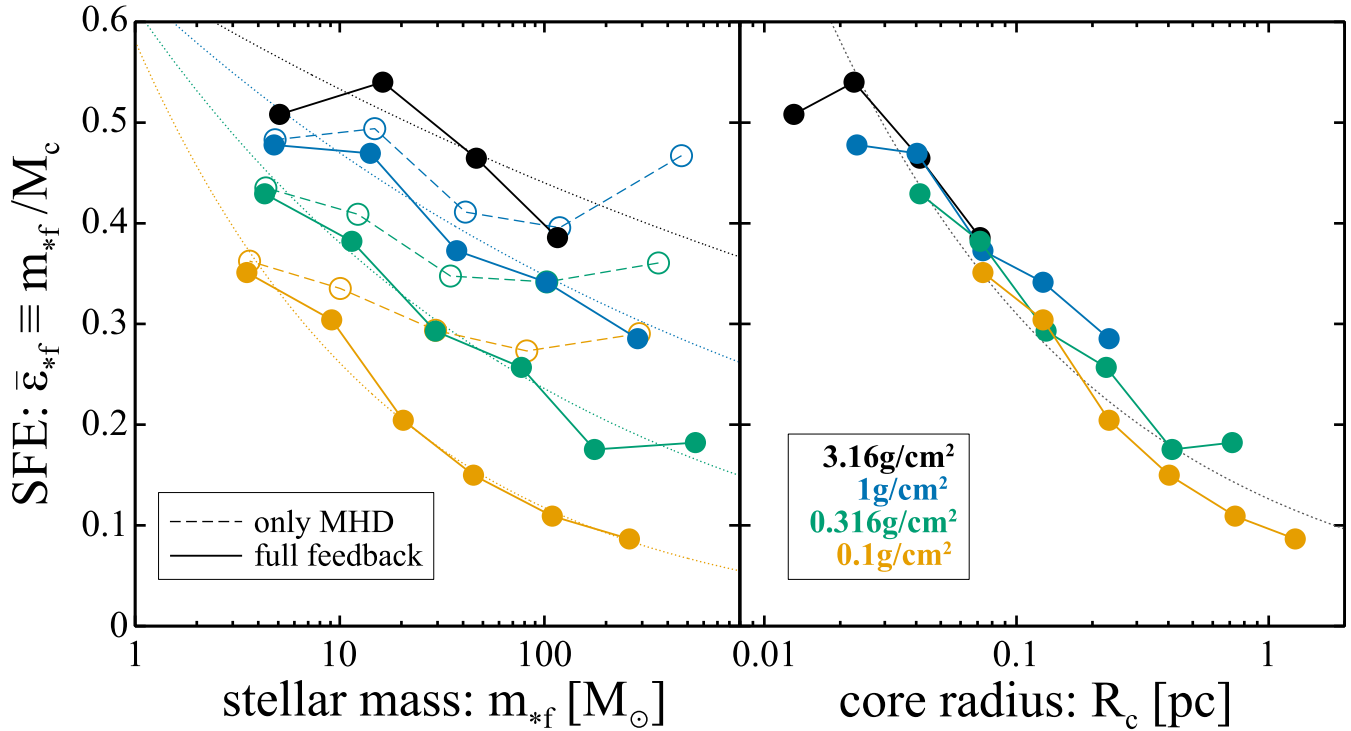}
\end{center}
\caption{
The SFE, $\sfe\tbond m_{\rm *f}/M_{c}$, for various models of
present-day massive (and intermediate-mass) star formation as
functions of final protostellar masses, $m_{\rm *f}$ ({\it left}), and
as functions initial core radii, $R_{\rm c}$ ({\it right}).  The solid
lines indicate SFEs from the fiducial full feedback models for cores
in different clump mass surface density environments, as labelled.
In the left panel, the dashed lines show SFEs evaluated with only MHD
disk wind feedback for comparison
(note that these models have not
been computed for the highest $\Sigma$ cases; see text).
The dotted lines show the fitting plots ({\it left:} eqs. \ref{eq_fit_a} and
\ref{eq_fit_b}, and {\it right:} eq. \ref{eq_fit2}).  }
\label{fig_sfe}
\end{figure*}

Now we return to the case of present-day massive star formation and
explore how the SFE, $\sfe$, depends on core mass, $M_{c}$, and
ambient clump mass surface density, $\Sigcl$.  The left panel of
Figure~\ref{fig_sfe} shows SFEs for $\Sigcl=0.1$--$3.16\:\gcm$ as
functions of final protostellar mass, $m_{\rm *f}$, i.e., when
accretion stops. The SFE with only MHD disk wind feedback has a weak
dependance on $m_{\rm *f}$, with values of $\sim0.3$--$0.5$, similar
to the results of \citet{mat00} and \citet{zha14}. 
Note, that for the highest $\Sigma$ case we have not run the MHD disk
wind only cases since their very high accretion rates lead to
protostellar structures that are difficult to model numerically with
our adopted protostellar evolution code.
On the other hand,
the SFE in the models with radiation feedback decreases quite strongly
with the final stellar mass for all $\Sigcl$ cases. The deviation from
the MHD disk wind only case is small if the final stellar mass is less
than $10\:\msun$.  The SFE becomes much smaller as $m_{\rm *f}$
increases, since radiative feedback grows strongly with stellar mass.
The results of $\Sigcl=0.1\:\gcm$ shows the strongest impact of
radiative feedback.  In this case, the SFE is only $0.1$ or less when
forming $\gtrsim100\:\msun$ stars.  On the other hand, for higher
$\Sigcl$, as we have seen in \S\ref{sec_evo}, the impact of radiative
feedback is smaller due to the higher accretion rates.  The dominant
feedback mechanism for setting SFEs is the MHD disk wind for
$\Sigcl\ga0.3\:\gcm$, even in the formation of very massive stars.

The right panel of Figure \ref{fig_sfe} shows the SFEs as functions of
initial core radii, $R_{c}$. We see that more compact cores result in
higher SFE.  Interestingly, all of our models with full feedback can
be fitted by a single power law of
\begin{eqnarray}
	\sfe\simeq0.31\left(\frac{R_{c}}{0.1{\rm pc}}\right)^{-0.39}, \label{eq_fit2}
\end{eqnarray}	
within an error of $35\%$.
This simple fitting formula is convenient analytic result that can be
applied as a sub-grid model to large scale simulations of star
formation that resolve formation of massive pre-stellar cores (note,
this result applies to cores from $10\:M_\odot$ to
$\sim10^3\:M_\odot$).

\section{Discussion} \label{sec_discussion}

First, we summarize the relative importance of different feedback
mechanisms in \S\ref{sec_relative}.  Then we discuss the potential
impact of radiation feedback on shaping the high-mass end of the IMF
in \S\ref{sec_impact}. Next we consider the metallicity dependence of
massive star formation in \S\ref{sec_metal}.  Finally, we note the
caveats and limitations of our modeling in \S\ref{sec_cav}.

\subsection{Relative importance of feedback processes}\label{sec_relative}

We have studied multiple feedback mechanisms, i.e., MHD disk wind,
radiation pressure, photoevaporation, and stellar wind, during star
formation via core accretion.  We find that for present-day massive
star formation at solar metallicity the MHD disk wind plays a dominant
role not only in low-mass star formation but also in massive star
formation.

In simple spherical core collapse radiation pressure acting on dusty
infall stops formation of massive star formation for
$m_*\gtrsim20\msun$.  However, in non-spherical disk accretion, the
optically thick inner region shields outer equatorial zone accretion.
Additionally, the MHD disk wind outflow cavity effectively reduces the
effects of radiation pressure by dust re-emission, i.e., $f_{\rm
  trap}\simeq1$. Using equations (\ref{eq_pdw}) and (\ref{eq_prad}),
we can compare the momentum injection rates from the MHD disk wind and
from radiation pressure:
\begin{eqnarray}
\frac{\dot{p}_{\rm rad}}{\dot{p}_{\rm dw}} \sim 0.03
\left(\frac{m_*}{50\msun} \right)^{1.8}
\left(\frac{\mdot_{\rm acc*}}{10^{-3}\msunyr} \right)^{-1}.
\end{eqnarray}
Here we use the luminosity and radius of the ZAMS model
\citep{sch92} and adopt $\phi_{\rm dw}\simeq0.2$ from our results
\citep[see also][]{zha14}. As seen from this evaluation, the MHD disk
wind momentum injection rate is much higher than that from radiation
pressure even for $m_*=100\:\msun$.  However, as described in
\S\ref{sec_outflow}, the MHD disk wind is collimated while the stellar
radiation acts isotropically.  Considering the angular distribution of
momenta (eq. \ref{eq_phidw}), we obtain the following relation,
\begin{eqnarray}
\left. \left.
\frac{d \dot{p}_{\rm rad}}{d \Omega} \right/
\frac{d \dot{p}_{\rm dw}}{d \Omega} \right|_{\theta=84\degr}
\simeq 0.2 \left(\frac{m_*}{50\:\msun} \right)^{1.8}  \nonumber \\
	\times \left(  \frac{\mdot_{\rm acc*}}{10^{-3}\:\msunyr} \right)^{-1}.
\end{eqnarray}
It can be seen that the component of radiation pressure is not
negligible at large angles $\theta$ away from the outflow axis.  Also,
these equations indicate that the contribution of radiation pressure
is higher at lower accretion rates, i.e., lower $\Sigcl$ cases.  The
accretion rate also becomes smaller when the outflow cavity opens up,
which enhances the importance of $p_{\rm rad}$. In this way, the MHD
disk wind supplies a large measure of momentum to create the outflow,
and the radiation pressure assists to open up the cavity and help set
the SFE.

As shown in \S\ref{sec_pe}, the photoevaporation mass-loss rate is
regulated by dust attenuation of ionizing photons and is a relatively
minor feedback process, unlike in the case of primordial star
formation. Of course, dust attenuation only occurs in the region where
dust survives, i.e., $r>r_{\rm sub}$. For this region, assuming a
constant recombination rate $\alpha_{\rm A}$, we can derive a simple
differential equation from equations (\ref{eq_n0_d}) and
(\ref{eq_taud}):
\begin{eqnarray}
\frac{d \uptau_{d}(r)}{dr}=\sigma_{\rm a,d} n_{\rm sub} x^{-1.5} e^{-\uptau_{d}/2},
\end{eqnarray}
where $n_{\rm sub}=n_{\rm 0}(r_{\rm sub})$ is the base density of the
photoevaporation flow at the dust sublimation front, and $x\equiv
r/r_{d}$ is a dimensionless radius. This equation has an analytic
solution of
\begin{eqnarray}
	\uptau_{d}(r)
	&=& 2\ln \left\{1 + \hat{\uptau}_{d}(1-x^{-0.5}) \right\}; \label{eq_tautau}\\
	n_0(r) &=& \frac{n_{\rm sub}x^{-1.5}}{1+\hat{\uptau}_{d}(1-x^{-0.5})}. \label{eq_n0_ana}
\end{eqnarray}
The characteristic optical depth of the system $\hat{\uptau}_{d}$,
which also appears in \S\ref{sec_pe}, is evaluated as
\begin{equation}
\hat{\uptau}_{d}\simeq100 \left(\frac{S_{\rm *acc}}{10^{50}\:{\rm s}^{-1}}\right)^{1/2}
\left(\frac{r_{\rm sub}}{30\:{\rm AU}} \right)^{-1/2}, \label{eq_cha_tau}
\end{equation}
from equation (\ref{eq_n0_df}) and assuming an ionized gas temperature
of $10^4\:{\rm K}$. This solution is consistent with the dust-free
case in the limit of $\hat{\uptau}_{d}\to0$. Considering limits of a
far distance of $x\gg1$, we see basic features of the effects of dust
attenuation, i.e., $\uptau_{d}\to2\ln(1+\hat{\uptau}_{d})$ and $n_{\rm
  0}\to n_{\rm sub}x^{-1.5}/(1+\hat{\uptau}_{d})$. The
photoevaporation flow reaches optically thick conditions, i.e.,
$\uptau_{d}=1$, when the characteristic optical depth reaches
$\hat{\uptau}_{d}\simeq0.7$. However, in the optically thick limit of
$\hat{\uptau}_{d}\gg1$, the total optical depth converges to
$2\ln\hat{\uptau}_{d}$, which is smaller than $\hat{\uptau}_{d}$.
Thus, the suppression of base density $n_0$ is not as strong as the
exponent of $e^{-\hat{\uptau}_{d}}$ and involves only a factor of
$\hat{\uptau}_{d}$. Due to this suppression of photoionization, the
total evaporation rate is also regulated to
$\Mdot_{\rm pe}/\Mdot_{{\rm pe},\uptau_{d}=0}\simeq1/\hat{\uptau}_{d}$
since the evaporation rate is proportional to $n_0$ (eq. \ref{eq_mdot_pe}).  In
this manner, dust attenuation of ionizing photons regulates the
mass-loss rate by photoevaporation when $\hat{\uptau}_{d}\gtrsim1$.

The stellar wind is found to be the weakest feedback process in our model. 
Please note that we have not explicitly considered the momentum
injection from the wind, since it is always sub-dominant compared to
the MHD disk wind and radiation pressure and would be confined along a
narrow region of the outflow axis.

The Eddington factor $\Gamma_{e}$ evaluated by our protostellar
calculation is typically higher than that of the ZAMS case, however it
is still smaller than the critical value of $0.7$ to initiate the
extreme wind mass-loss regime (Fig. \ref{fig_sw}). Moreover, even
assuming the maximum Eddington factor of $\Gamma_{e}=1$ in equation
(\ref{eq_mdot_sw}), the stellar wind mass-loss rate is lower than
$10^{-4}\msunyr$ at $100\msun$.  Therefore, we conclude that, the
stellar wind is not important during the protostellar accretion phase.
Note, however, that during evolution after the mass accretion phase
over timescales of $\rm \sim Myr$,
the stellar wind has important effect leading to significant mass-loss.

To conclude, in massive star formation by core accretion, we find that
the MHD disk wind is most important feedback mechanism, radiation
pressure assists the opening-up of the outflow cavity to wide angles,
photoevaporation is regulated by the dust attenuation and is thus of
minor importance, and stellar wind mass-loss has a very minor effect
during the accretion phase. In the sense that the bipolar MHD-driven
outflow is the most significant feedback, massive star formation
resembles low-mass star formation, however SFEs can be significantly
reduced by the action of radiative feedback.

\subsection{Implications for the high-mass end of the IMF}\label{sec_impact}

As we have seen, radiative feedback can significantly reduce SFE for
the formation of very massive stars. Considering the stellar IMF to be
the result of a multiplicative combination of the CMF and SFE, we can
expect that the effects of radiative feedback may be seen in the
high-mass end of the IMF. While the MHD disk wind only feedback sets a
SFE, which depends only weakly on the initial core mass, the full
model including radiative feedback yields smaller SFEs for higher core
masses, $M_{c}$ (Fig. \ref{fig_sfe}). Potentially this could induce a
steepening of the IMF at the highest masses and if this is steep
enough it may appear as an apparent truncation.
Using the obtained SFE, we can relate the IMF to the CMF. We introduce
the exponent of $\varepsilon' \equiv {d \ln \sfe}/{d \ln M_{c}}$.
If the CMF is a power-law distribution of $d\N/d \ln M_{c} \propto
M_{c}^{-\alpha_{c}}$ and the exponent of $\varepsilon'$ is
constant, then the IMF would be
\begin{equation}
\frac{d\N}{d \ln m_{\rm *f}}\propto m_{\rm *f}^{-\alpha_{c}/(1+\varepsilon')},
\end{equation}
\citep{nak95,mat00}. The SFEs for massive cores with
$M_{c}=10$--$3000\:\msun$ and $\Sigcl=0.1$--$3.16\:\gcm$ obtained by
our model are well fitted by
\begin{eqnarray}
\sfe &\simeq& 0.668 \left( \frac{M_{c}}{\msun} \right)^{\varepsilon'}, \label{eq_fit_a}\\
\varepsilon' &\simeq&-0.115 \left( \frac{\Sigcl}{\gcm} \right)^{-0.35}, \label{eq_fit_b}
\end{eqnarray}
within an error of  $15\%$.
Then, the power law exponent of the IMF at $>10\msun$ is estimated as
$-1.13\alpha_{c}$ in clumps with $\Sigcl=1\:\gcm$.  Assuming an
initial CMF slope of $\alpha_{c}=2.35$, i.e., the same as the Salpeter
IMF \citep{sal55}, the total number of stars with $10$--$100\:\msun$
is smaller than than that simply expected from the CMF slope by about
$11(\Sigcl/\gcm)^{-0.35}\%$ due to MHD disk wind and radiative
feedback. However, this reduction of massive stars is too small 
(the reduction factor is about 55\% for the mass range
100--$300\:M_\odot$) to explain the cut-off at $150\:\msun$ reported
for the Arches cluster \citep{fig05}. Thus we conclude that the
high-mass end of IMF, especially its potential truncation at masses
$\sim150$--$300\:M_\odot$, is mainly determined by the pre-stellar
core mass function rather than by feedback.

\subsection{Radiation feedback in massive star formation at different metallicities}\label{sec_metal}

In this paper, we mainly study the formation of massive stars at solar
metallicity. However, we have also applied our model to the primordial
star formation case (\S\ref{sec_popIII_model} \&
\ref{sec_popIII_results}) in order to compare to previous studies and
to obtain a basic insight into the effect of metallicity. Our results
show that the impact of radiative feedback depends strongly on such
metallicity changes. Since massive stars are thought to have been
important thoughout cosmic history as metallicities have evolved from
primoridal, near-zero limits to $\sim$solar values and beyond,
here we give some general discussion about the dependence of radiative
feedback, especially radiation pressure and photoevaporation mass-loss
(stellar wind feedback is weak at solar metallicity and would be even
weaker at lower metallicities). However, we defer a detailed
quantitative investigation of massive star formation at intermediate
metallicities of $0<Z<Z_{\odot}$ to a future paper.

Radiation pressure is the strongest radiative feedback mechanism at
solar metallicities.  However, since it acts on dust grains in the
infalling envelope, it must depend on metallicity. We have ignored
dust re-emission since it is assumed to effectively escape from the
outflow cavities.
Then, the momentum injection by radiation pressure can be evaluated
assuming $f_{\rm trap}=1$ in equation (\ref{eq_prad}), as long as the
envelope is optically thick for direct stellar radiation.  In other
words, the effect of radiation pressure becomes weaker if the
metallicity is low enough to keep the envelope transparent for direct
stellar radiation. This transparency depends on the stellar spectrum
and the grain components, but the typical opacity for direct stellar
radiation is approximately evaluated as
$\kappa_*\sim100(Z/Z_\odot)\:{\rm cm^{2}\:g^{-1}}$ for massive stars
assuming the opacity is simply proportional to metallicity.  Then, the
trapping factor is approximately given by
\begin{equation}
	f_{\rm trap} \sim \min\left[1, 100 \left(\frac{Z}{Z_\odot}\right)
	\left( \frac{\Sigcl}{\gcm} \right)
	\right].
\end{equation}
Assuming a typical massive core always forms in a clump with a mass
surface density of $1\:\gcm$, the effect of radiation pressure would
become weaker for metallicities of $Z\la 10^{-2}\zsun$.

Photoevaporation is strongly suppressed at solar metallicities because
of the dust attenuation of ionizing photons.  As described
\S\ref{sec_relative}, the photoevaporation rate with dust attenuation
is about $1/\hat{\uptau}_{d}$ of that of the dust-free case if
$\hat{\uptau}_{d}\gg1$.  Then, if the dust opacity for ionizing
photons is simply proportional to the metallicity, we obtain the
following relation of
\begin{equation}
\frac{\Mdot_{\rm pe}}{\Mdot_{{\rm pe},\uptau_{d}=0}} \sim \frac{1}{1+100(Z/\zsun)}.
\end{equation}
Here we assume $\hat{\uptau}_{d}\sim100$ at solar metallicity as a
typical value for the high ionizing photon production rate of
$10^{50}\:{\rm s}^{-1}$ at which photoevaporation mass-loss rate could
be important (eq. \ref{eq_cha_tau}). Therefore, the photoevaporation
mass-loss rate could be as high as that at zero metallicity at
metallicities of $Z\la10^{-2}\zsun$.

Considering both radiation pressure and photoevaporation, the critical
metallicity for their transitions coincide at $\sim10^{-2}\zsun$.
Dust absorption is efficient at higher metallicities than this
critical value, which means that radiation pressure acts
effectively. On the other hand, photoevaporation is suppressed at
these higher metallicities. In the lower metallicity regime dust
absorption is weak, which lessens the impact of radiation pressure,
while photoevapration is more effective. These considerations suggest
that the total effects of radiative feedback may be strongest at
$\sim10^{-2}\zsun$.

However, note that we are not suggesting that massive star formation
is necessarily rarer at metallicities of $\sim10^{-2}\:\zsun$.
Only that SFE could be lower.
The core mass function will also play an important role,
along with the typical clump mass surface density.
It is difficult to predict how these will vary with metallicity,
especially since they may also be more strongly influenced by the
degree of magnetization of the gas.
Other processes, such as disk fragmentation \citep[e.g.][]{TO14}, may also play a role.

\subsection{Caveats}\label{sec_cav}

Even though our model predictions, including those from previous
papers in this series, have some agreements with observations
\citep{zha13a,tan16}, this is a semi-analytic model that is still
highly simplified and idealized.
Ultimately, the predictions of the model need to be tested by full
radiation-MHD numerical simulations, especially to study the
interaction of the outflow with the infall envelope and establishment
of the outflow cavity boundary. Below we discuss some additional
caveats of our modeling.

We have considered only single star formation. The massive cores are
expected to be supported mainly by non-thermal pressures, i.e.,
turbulence and magnetic fields, which keeps them from fragmenting to
the thermal Jeans mass of $\sim1\msun$ \citep[e.g.,][]{mck03}.  Also
the catastrophic fragmentation during collapse is expected to be
suppressed by radiative heating by the high accretion luminosity and
the efficient angular momentum transportation by magnetic breaking
\citep{kru07,com11}.  However, a small amount of fragmentation may
still occur, as seen simulations such as \citet{kru09}.  Indeed, it is
observationally known that more than $70\%$ of massive stars have
close companions which eventually exchange their masses \citep{san12}.
We expect that our model is still quantitatively appropriate since the
feedback is dominated by a single object as long as the total stellar
mass is dominated by a primary star in the binary/multiple system.

On the other hand, our feedback model would need significantly
modification if the system contains similar mass stars. Qualitatively,
we expect that radiative feedback in the case of similar mass binaries
would be weaker than that in the case of formation of a single massive
star. This is because the stellar luminosity increases nonlinearly
with the mass. If some amount of material is divided into two objects,
the total luminosity is smaller than that of a single star with the
same total mass.  In contrast, the momentum rate from MHD disk winds
is roughly proportional to the total accretion rate.  Therefore, we
expect that the conclusion that the MHD disk wind is the dominant
feedback is correct also in the case of formation of a massive binary.

Next, we did not study the case with very high accretion rate of
$\mdot_*\ga10^{-2}\:\msunyr$, which would arise in the collapse of
very unstable cores of $M_{c}\ga1500(\Sigcl/\gcm)^{-1}\:\msun$
(eq. \ref{eq_Mdot}).  Even though the typical accretion rate of
massive star formation is thought to be of the order of
$10^{-3}\:\msunyr$, there may be cases with higher accretion rates
that are especially important for formation of very massive stars.  In
the case of zero metallicity, \citet{hos12} have found a new branch of
protostellar evolution at $\ga10^{-2}\msunyr$: the protostar balloons
as $r_*\simeq2.6\times10^3(m_*/100\:\msun)^{1/2}\:\rsun$ without KH
contraction.  Such ``supergiant'' protostars have low-effective
temperatures of about $5000\:{\rm K}$ and thus photoevaporation
becomes ineffective, which is normally the most important feedback at
zero metallicity. A similar phenomenon may appear also at solar
metallicities, but it is non-trivial to calculate this evolution
because of the presence of metals that alter the protostellar
evolution \citep{hos09,hos09b}. Due to this uncertainty of
protostellar evolution and also the numerical difficulty of
calculation of supergiant protostars, we have avoided models with
parameter range of $\ga10^{-2}\:\msunyr$ in this paper.  However, even
without detailed calculations, we can expect that the MHD disk wind is
still the most dominant feedback in the cases of such rapid accretion.
This is because the momentum rate of MHD disk wind is proportional to
the accretion rate, while the radiation pressure acts similarly as in
lower-accretion rate case, and the photoevaporation becomes negligible
in the supergiant protostar phase if it appears at solar metallicity.
Future work on accurate protostellar evolution calculations with
$\ga10^{-2}\msunyr$ at solar metallicities is needed to confirm this
expectation.

We also did not consider short timescale variations of accretion
rates, which may be induced by disk instabilties, e.g., due to
self-gravity.  \citet{mey17} simulated the formation of a massive star
and showed the accretion bursts occur repetitively.  The accretion
rate rapidly increases from $10^{-4}$--$10^{-3}\msunyr$ to
$10^{-1}\msunyr$ within a duration of $10$ years and it recurs with
the timescale of several kyr.  The accretion burst has a significant
impact on the observational aspects, since it results in luminosity
outbursts similar to FU Orionis objects.  Considering the evolution of
the infalling envelope under the influence of feedback, however, we
suspect that the accretion burst does not have too significant an
impact.  This is because the global evolution of the infalling
envelope is affected by the accretion rate averaged over the accretion
timescale of
$\sim10^4(m_*/10\msun)(\mdot_*/10^{-3}\msunyr){\rm~yr}$, which is
longer than the expected durations and recurrence timescales of
accretion bursts. Accretion bursts would also change the protostellar
evolution, since, as described in the previous paragraph, such
high-accretion rates can cause a supergiant phase.  However,
\citet{sak15} showed that, at least in zero-metallicity case, the
supergiant phase cannot last as long as the recurrence timescale of
$\ge10^3~{\rm yr}$ since the KH timescale is very short.  To check
these speculations, further study, deferred to a future paper, is
needed to include accretion bursts self-consistently in our modeling.

In addition to further theoretical and numerical studies, better
observational tests are needed to confirm the reliability of our
theoretical model. We have applied the previous versions of our model
to make predictions of observational features using radiative transfer
calculations \citep{zha11,zha13,zha14,tan16}. In a future paper we
will model the radiative transfer predictions of the feedback models
that we have presented here.

%%%%%%%%%% Section 5 %%%%%%%%%%
\section{Conclusion}\label{sec_conclusion}

We have investigated the impact of multiple feedback mechanisms in
massive star formation by core accretion, and calculated the star
formation efficiency (SFE) from pre-stellar cores. Our model includes
feedback by the outflows driven by momenta from MHD disk winds and
radiation pressure, and the effects of mass-loss by photoevaporation
and stellar winds.  We found the MHD disk wind is the dominant
feedback mechanism for all cases considered, while radiation pressure
can cause a significant reduction in SFE at the highest masses and
especially in lower mass surface density clumps.
The obtained SFE can be fitted as $0.4(M_{c}/100\:\msun)^{-0.115}$ in
the initial core mass range of $M_{c}=10$--$1000\:\msun$ at the
ambient clump mass surface density of $1\:\gcm$, which is a typical
value for massive star formation.  The gentle decline of
$M_{c}^{-0.115}$ is caused by the radiative feedback which is stronger
at higher masses.  Therefore, we conclude that the shape of high-mass
end of initial stellar mass function, especially potential truncation
at $m_*\sim150$--$300\:M_\odot$, is mainly determined by the
pre-stellar core mass function and/or disk fragmentation rather than
the effects of feedback.

The MHD disk wind provides the major portion ($\ga90\%$) of outflow
momentum over all the considered mass range, and drives the outflow
before the stellar mass reaches about $20\msun$, when radiation
pressure acting on dust grains in a spherical envelope becomes
stronger than the gravitational force. Such radiation pressure was
once thought to be a potential barrier for massive star formation, but
in more realistic disk accretion and outflow cavity geometries, the
strong direct stellar radiation is shielded in the disk-shadowed
region, and dust re-emission escapes along the cavities.  Therefore,
feedback by radiation pressure is not a catastrophic problem for
massive star formation. Still, although the total momentum is
dominated by the MHD disk wind, radiation pressure also assists to
open the outflow cavity to wider angles, since it acts more
isotropically than the collimated MHD outflow.

Mass-loss by photoevaporation is strongly suppressed by dust
attenuation of ionizing photons. When the protostar starts the
Kelvin-Helmholtz contraction at $10$--$20\:\msun$, the ionizing photon
rate increases with the effective temperature and the photoevaporation
starts.  However, as the mass-loss rate increases, the
photoevaporation flow becomes opaque for the ionizing radiation due to
dust opacity. Thus, the photoevaporation mass-loss rate is regulated
to $\la10^{-4}\msunyr$, which is $<10\%$ of the mass-loss rate without
dust attenuation.  Since the typical accretion rates of massive star
formation are $10^{-3}\msunyr$, photoevaporation has only a minor
impact for the SFE at solar metallicities.

The mass-loss by stellar wind is found to be $\la10^{-5}\msunyr$, and
thus is not important to set the stellar mass.  The stellar wind is,
however, very important for the later evolution over timescales of
several million years, i.e., the total mass-loss can be tens of
$\msun$.

We also applied our model to primordial, Pop III star formation.  Due
to the lack of dust grains at zero metallicity, the radiation pressure
is negligible and also dust attenuation of ionizing photons does not
occur.  Therefore, photoevaporation is the major feedback effect in
primordial star formation.  In our fiducial model, the
photoevaporation rate reaches $\sim10^{-3}\:\msunyr$ and stops the
mass accretion at about $44\msun$. In this
manner, radiation feedback depends on metallicity, mainly due to the
dust absorption.  We evaluated that the critical metallicity for two
radiative feedback transitions is $\sim10^{-2}\zsun$. Dust absorption
is effective at higher metallicities than this critical metallicity,
which results in radiation pressure being strong while
photoevaporation is suppressed.  On the other hand, at lower
metallicities dust absorption is weak and so radiation pressure
eventually becomes negligible and the photoevaporation is more
important. Since massive stars are thought to have been
astrophysically important since the times of the first stars, more
detailed studies are needed to investigate the quantitative effects of
feedback as a function of metallicity.

%%%%%%%%%%%%%%%%%%
\section*{Acknowledgments}
The authors thank Jorick Vink, Jan E. Staff, Takashi Hosokawa,
Kohei Inayoshi, and Masanobu Kunitomo for fruitful discussions and comments.
JCT acknowledges support from NSF grants AST 1212089 and AST 1411527.

%\appendix

\clearpage

\end{document}